\documentclass{aa}  

\usepackage{graphicx}
\usepackage{txfonts}
\usepackage{xcolor}

\def\la{\mathrel{\hbox{\rlap{\hbox{\lower4pt\hbox{$\sim$}}}{\raise2pt\hbox{$<$}}}}}
\def\ga{\mathrel{\hbox{\rlap{\hbox{\lower4pt\hbox{$\sim$}}}{\raise2pt\hbox{$>$}}}}}

\begin{document}

   \title{Evolution of the truncated disc and inner hot-flow of GX~339--4}


   \author{P. Chainakun
          \inst{1}\fnmsep\thanks{pchainakun@g.sut.ac.th}
           \and
          W. Luangtip\inst{2,3}
          \and
          A. J. Young\inst{4}
           \and
          P. Thongkonsing\inst{1}
           \and
          M. Srichok\inst{1}
          }

   \institute{School of Physics, Institute of Science, Suranaree University of Technology, Nakhon Ratchasima 30000, Thailand\\
              \email{pchainakun@g.sut.ac.th}
         \and
            Department of Physics, Faculty of Science, Srinakharinwirot University, Bangkok 10110, Thailand
          \and
          National Astronomical Research Institute of Thailand, Chiang Mai 50180, Thailand
          \and
          H.H. Wills Physics Laboratory, Tyndall Avenue, Bristol BS8 1TL, UK
             }

   \date{Received xxx; accepted yyy}

 
  \abstract
   {}
   {We study the changes in geometry of the truncated disc and the inner hot-flow of GX~339--4 by analyzing the Power Spectral Density (PSD) extracted from six {\it XMM-Newton} observations taken at the very end of an outburst.}
   {A theoretical model of the PSD of GX~339--4 in the 0.3--0.7 keV (thermal reverberation dominated) and 0.7--1.5 keV (disc continuum dominated) energy bands is developed. The model assumes the standard accretion disc to be truncated at a specific radius, inside of which are two distinct hot-flow zones: one spectrally soft and the other spectrally hard. The effects of disc-fluctuations and thermal reverberation are taken into account. }
   {This model successfully produces the traditional bumpy PSD profiles and provides good fits to the GX~339--4 data. The truncation radius is found to increase from $r_{\rm trc} \sim10$ to $55r_{\rm g}$ as the source luminosity decreases, strongly confirming that the truncation radius can be characterized as a function of luminosity. Keeping in mind the large uncertainty in previous measurements of the truncation radius, our values are larger than some obtained from spectroscopic analysis, but smaller than those implied from reverberation lag analysis. Furthermore, the size of two inner hot-flow zones which are spectrally hard and spectrally soft are also growing from $\sim 5$ to $27r_{\rm g}$ and from $\sim 3$ to $26r_{\rm g}$, respectively, as the flux decreases. We find that the radial range of inner hard zone is always larger than the range of the soft hot-flow zone, but by a comparatively small factor of $\sim 1.1$--$2.2$}
   {}


   \maketitle
%

\section{Introduction}

The X-ray variability in both AGN and X-ray binaries could be assessed through the use of the Power Spectral Density (PSD). The PSD describes how the X-ray power varies on different timescales (or different temporal frequencies), which depends on the mechanisms responsible for the X-ray production near black holes \cite[e.g.,][]{Martin2012}. Another timing technique to probe the innermost regions is by measuring X-ray reverberation lags, i.e., the time delays between the changes in direct continuum and reprocessing echoes from the disc \citep[see][for a review]{Uttley2014}. Due to the longer distance travelled by the reflected photons, changes in energy bands dominated by reflection, or reprocessing, lag behind changes in the continuum dominated band. The first hint of such reverberation delays was reported by \cite{McHardy2007} in the AGN Ark 564, following by the first robust detection by \cite{Fabian2009} in the AGN 1H0707--495. 

X-ray reverberation in black hole X-ray binaries was first robustly detected in GX~339--4 by \cite{Uttley2011} when the source was in its hard state. Previous studies of GX~339--4 pointed to the approximate central mass being $\geq 6 M_{\odot}$ \citep[e.g.,][]{Hynes2003} and a small disc inclination angle \citep{Demarco2015a}. \cite{Miller2008} fit the {\it Suzaku} spectra and found the central black hole has a very high spin $a \sim 0.998$. The X-ray spectroscopic analysis of the hard state spectra from the {\it RXTE} archive carried out by \cite{Garcia2015} suggested the black hole spin to be $a \sim 0.95$. Spectral fitting of GX~339--4 during its very high flux state using $\emph{NuSTAR}$ and $\emph{Swift}$ also suggested a high spin of $a \sim 0.95$ \citep{Parker2016}. According to the time-lag analysis, \cite{Uttley2011} found that the disc thermal emission ($\sim 0.3$--0.7~keV, soft band) leads the power-law variations ($\sim 0.7$--1.5~keV, hard band) on long timescales ($>1$s). \cite{Mahmoud2019} assumed the soft component that leads the power-law emission is a soft Comptontized component. \cite{Rapisarda2016, Rapisarda2017} instead modelled it as a variable inner region of the thin disc. However, the disc blackbody variations lag behind the power-law variations by a few milliseconds on short timescales ($< 1$s). This switch from low-frequency hard to high-frequency soft lags is thought to be produced by two distinct mechanisms. While the hard lags are likely due to inward propagating fluctuations \citep[e.g.,][]{Kotov2001,Arevalo2006}, the soft lags can be explained by thermal reverberation associated with the longer light-travel time the hard photons take from the central power-law X-ray source to the disc where they are reprocessed into relatively soft blackbody emission. The thermal reverberation lags then provide clues to the geometry of the X-ray source and the inner accretion flow close to the event horizon of the central black hole.   

The timing analysis of GX~339--4 including additional {\it XMM-Newton} observations revealed that the changes in time lags in the high flux states can be characterized as a function of luminosity \citep{Demarco2015a}. The luminosity may change with the truncation radius. When the luminosity increases, the truncation radius becomes smaller so that the observed flux increases due to additional disc dissipation and reflection from the inner disc. Meanwhile, the amplitude of reverberation lags become smaller as the photons reflecting off the inner radii of the disc have a smaller average light crossing time. On the other hand, if the truncation radius increases, the luminosity decreases and the reverberation lag increases. \cite{Reig2018} demonstrated that the correlation between the average time lag and the photon index (i.e., the lags increase as the X-ray continuum becomes softer) in the black-hole X-ray binaries could be a result of inverse Comptonization in the base of the jet. \cite{Kylafis2018} also reported this tight correlation in GX~339--4. \cite{Sridhar2020} found that the temperatures of both inner disc and corona are sensitive to the luminosity of state transition. 

The physical properties as well as the geometry of the Comptonizing region in GX~339--4, however, are still unclear. Different fitting techniques also lead to different implied values of truncation radius. \cite{Wang2018} investigated the mean spectrum of this source during a failed outburst in 2013 observed by {\it NuSTAR} and {\it Swift}. They found that a maximum truncation radius could reach $\sim 37r_{\rm g}$ ($1r_{\rm g}=1GM/c^{2}$ is the gravitational radius, where $G$ is the gravitational constant, $M$ is the black hole mass, and $c$ is the speed of light). They also reported a smaller truncation radius of $\sim 3$--$15r_{\rm g}$ during the 2015 outburst. Evolution of the reverberation lag at the end of the 2015 outburst and during the return to quiescence was investigated by \cite{Demarco2017}. By converting the amplitude of the lags to light-crossing distance, \cite{Demarco2017} estimated the truncation radius to be $\sim40$--$200r_{\rm g}$. \cite{Kylafis2018} showed that during the hard and hard-intermediate states the inner disc could extend inwards shrinking the hot inner flow and the jet base. \cite{Garcia2019} carried out a spectroscopic analysis of this source when it went through the failed outburst in 2017. They found that a dual-lamppost model provides a better fits than the standard single lamppost source and implied the truncation radius to reach a few gravitational radii as the luminosity increases in the hard state. 

While some of the spectroscopic studies of the refection component in the bright hard state suggested that the inner disc could extend very close to the innermost stable circular orbit  
\citep[e.g.][]{Garcia2015, Steiner2017, Wang2018}, there is also previous literature that reported the significantly larger truncation radius of, e.g., $>35r_{\rm g}$ \citep{Tomsick2009} during the hard state. The truncation of the optically thick, geometrically thin disc in the hard state can be the result of thermal conduction of heat from the corona that causes the inner disc to evaporate \citep{Meyer2000}. \cite{Mahmoud2019} modelled the hard state spectral-timing data of GX~339--4 taking into account the time lags due to reverberation and found evidence of the truncation radius in the order of $\sim20r_{\rm g}$. 

\cite{Veledina2016, Veledina2018} investigated the interference of X-rays under the propagating-fluctuations framework and found that the presence of multiple sources (e.g., disc Comptonization and synchrotron Comptonization sources) could produce a complex PSD shape seen in X-ray binaries. Reverberation signatures can also be imprinted the the PSD profiles \citep{Papadakis2016, Chainakun2019a}. The aim of our work is to develop a PSD model, taking into account both propagating-fluctuations and reverberation effects, to predict the changes in the accretion disc and the inner hot flows of GX~339--4. It is clear that there is a large scatter in the reported values of the truncation radius in GX~339--4 \citep[see, e.g.,][and references therein]{Wang2018}. Here we focus on the very last stages of the soft-to-hard transition as the source returns to quiescence, similar to the observations analyzed by \cite{Demarco2017}. 

The GX~339--4 observations used here and the data reduction are explained in Section 2. The theoretical PSD models are presented in Section 3. The fitting results are shown in Section 4, followed by the discussion in Section 5. The conclusions are drawn in Section 6.

\section{Observations and data reduction}

The data analysed in this paper were selected from those that were in the final phase of GX~339--4 outburst observed during August--September 2015, and were obtained from {\it XMM-Newton} Science Archive.\footnote{\url{http://nxsa.esac.esa.int/}} The selected observations are listed in Table~\ref{tab:xmm_obs}. All observational data were cleaned following the standard data reduction method as described on the {\it XMM-Newton} data analysis webpage{\footnote{\url{https://www.cosmos.esa.int/web/xmm-newton/sas-threads}}} using Science Analysis Software (SAS) version 18.0.0 with the latest calibration files. We then created EPIC-pn light curves using the SAS task {\sc evselect} with the selection criteria of PATTERN $\leqslant$4 and the time bin size of 6 ms. The source extraction region for the data in timing mode was defined as the data which has 28 $\leqslant$ RAWX $\leqslant$ 48 while the region for those observed in small window mode was defined as the circular area centred on the source position with radius of $40''$. 

Since the  source is very bright, the light curves should be dominated by the source counts. Here, we followed the data reduction outlined in \cite{Demarco2017}. To maximise the data signal to noise ratio, we did not remove background flaring events from the observational data. In fact, the fraction of the useful exposure time affected by the flaring events for each observation is $\la$ 7\%, excepting for the observation O1 in which $\sim60$ \% of the exposure time is affected. Nevertheless, we checked that timing products, i.e. the power spectra, obtained from the observation with and without removing the flaring events are well consistent. This verifies that the flares did not affect the GX~339--4 timing properties in our a timing analysis. We also note here that all observations were not affected by pile-up. In fact, the source count rates during the observations O1 - O4 are substantially below the maximum count rate allowed for an observation in timing mode of the pn detector ($<$240 count s$^{-1}$). For the observations O5 - O6 which were observed in small window mode, we ran the SAS task {\sc epatplot} to check and confirm that the observations were not significantly affected by the pile-up.  Therefore, the useful exposure time for each observation after performing the data reduction is shown in column 4 of Table~\ref{tab:xmm_obs}. The light curves in two different energy bands -- 0.3--0.7 keV and 0.7--1.5 keV -- were extracted from all pn observations. These bands are dominated by thermal reverberation and power-law continuum, respectively, as previously reported by \citet{Demarco2017}.

Finally, based on the light curves obtained, we created the power spectra using the {\sc ftools} task {\sc powspec}.\footnote{\url{https://heasarc.gsfc.nasa.gov/lheasoft/ftools/fhelp/powspec.txt}} In brief, each light curve was divided into a number of segments with the interval of 41 s and the time bin size of 6 ms; then these segments were converted into power spectra and averaged to create the single power spectrum for each light curve. The power spectra were rebinned by a factor of 1.04 dex for the data below 2 Hz and 1.2 dex for the data above 2 Hz. We ignored data points consistent with zero (or negative) power. The power spectra obtained from this method were then used as the basis for further analysis.

\begin{table}
\begin{center}
   \caption{\emph{XMM-Newton} observations of GX~339--4.} \label{tab:xmm_obs}
    \begin{tabular}{lccc}
    \hline
    Obs. ID & Date& Mode$^{a}$ & Exposure$^{b}$ \\
    & &  &  (ks) \\
    \hline
     0760646201 (O1)     & 2015-08-28  & Timing & 14.7 \\
     0760646301 (O2)     & 2015-09-02  & Timing & 15.7 \\
     0760646401 (O3)     & 2015-09-07  & Timing & 20.2 \\
     0760646501 (O4)     & 2015-09-12  & Timing & 18.6 \\
     0760646601 (O5)     & 2015-09-17  & Small Window & 36.5 \\
     0760646701 (O6)     & 2015-09-30  & Small Window & 33.4 \\
     \hline
     \multicolumn{4}{l}{\footnotesize \textit{Note.} $^{a}${X-ray instrument operating mode.} $^{b}${Useful exposure}} \\
     \multicolumn{4}{l}{\footnotesize {time after data cleaning.}} \\
     \end{tabular}
\end{center}
\end{table}

\section{Theoretical model}

\subsection{Geometry setup}

\begin{figure}
    \centerline{
        \includegraphics[width=0.45\textwidth]{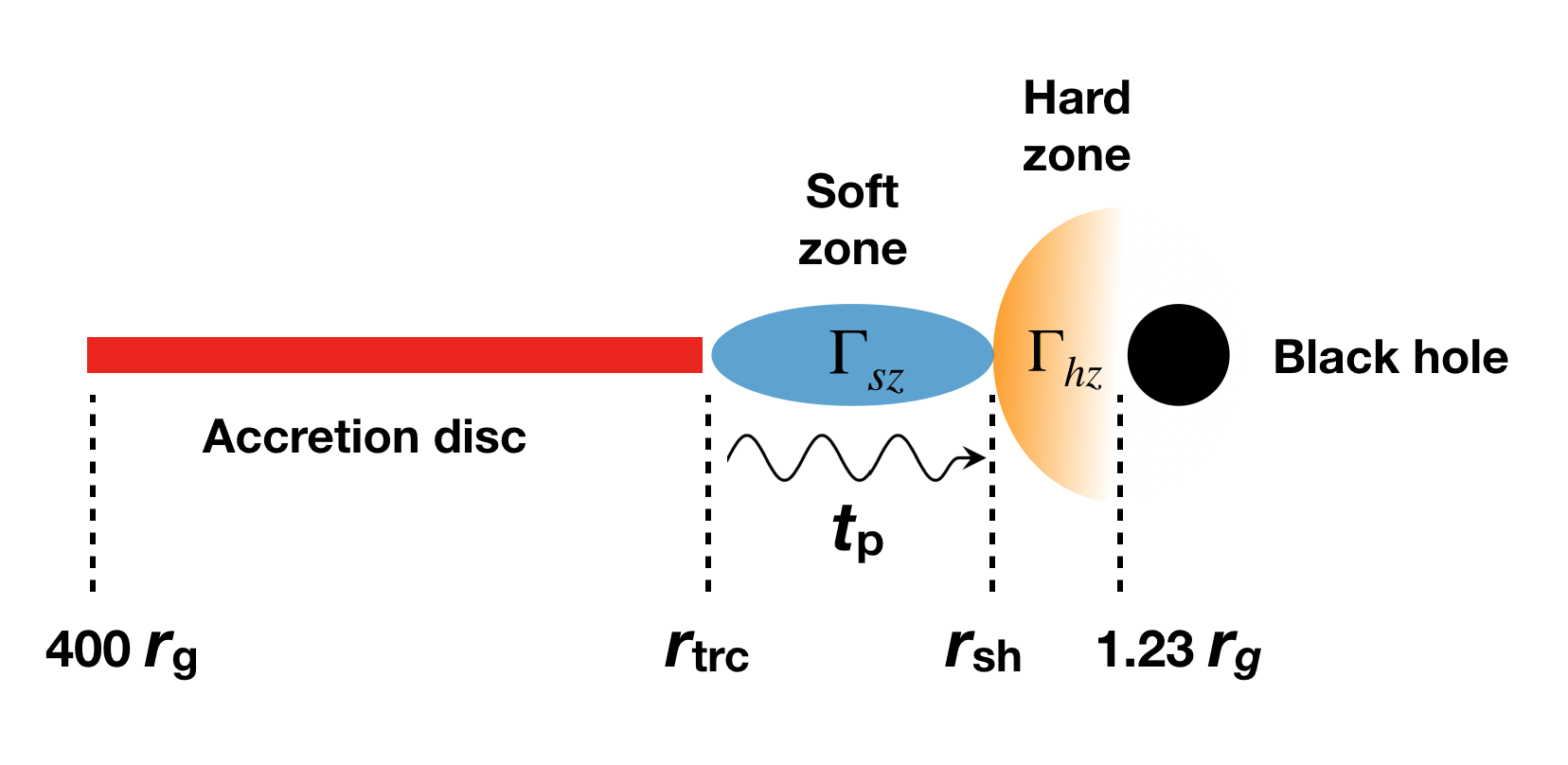}
    }
    \caption{The geometry of our model. The accretion disc varies on long timescales and extends from $400r_{\rm g}$--$r_{\rm trc}$. The inner regions from $r_{\rm trc}$--$r_{\rm sh}$ and $r_{\rm sh}$--$1.23r_{\rm g}$ are turbulent inner flows varying intrinsically on fast timescales which are responsible for the soft and hard spectral emission, respectively. The accretion disc provides seed photons to the inner hot flows, and reprocess and reflect the Comptonized emission photons coming back from the inner regions. The parameter $t_{\rm p}$ is the time taken for fluctuations to propagate from $r_{\rm trc}$ to $r_{\rm sh}$. 
    }
    \label{geometry}
\end{figure}

To avoid having too many free parameters, we fix the black hole mass of GX~339--4 to be $10M_{\odot}$, the inclination angle $i=30^{\circ}$, and the black hole spin $a=0.998$ \citep[e.g.,][]{Hynes2003, Demarco2015a,Garcia2015,Parker2016}. Therefore, the innermost stable circular orbit (ISCO) is at $\sim 1.23r_{\rm g}$. The disc is geometrically thin and optically thick \citep{Shakura1973} ranging between $400r_{\rm g}$--$r_{\rm trc}$, where $r_{\rm trc}$ is the truncation radius. Inside $r_{\rm trc}$ the disc is replaced by two hot-flow zones responsible for the emission of soft spectrum ($\Gamma_{\rm sz}$) and hard spectrum ($\Gamma_{\rm hz}$) ranging between $r_{\rm trc}$--$r_{\rm sh}$ and $r_{\rm sh}$--$1.23r_{\rm g}$, respectively. The $\Gamma_{\rm sz}$ and $\Gamma_{\rm hz}$ are the photon indices of the X-ray continuum emitted from the inner soft and hard zones, respectively, and $r_{\rm sh}$ is the transition radius between the two zones. The accretion disc varies on long timescales while the turbulent inner hot-flows vary intrinsically on relatively fast timescales. The parameter $t_{\rm p}$ characterises the time taken for fluctuations to propagate from $r_{\rm trc}$ to $r_{\rm sh}$. Our geometric setup is presented in Fig~\ref{geometry}. 

We consider two distinct mechanisms that take place in our system: fluctuations in the mass accretion rate and the thermal reverberation. The accretion disc provides seed photons to the inner hot-flow zones, and reflects the Comptonised photons travelling outwards from the hot-flows. The continuum flux in the energy band of interest $E_{j}$ emitted from the spectrally soft and the hard hot-flow zones are
\begin{eqnarray}
    F_{\rm sz}(E_{j}) & \propto & \int_{E_{j,\rm low}}^{E_{j,\rm high}} {E_{j}^\prime}^{-\Gamma_{\rm sz}} {\rm d} E_{j}^\prime \;, \\
    \label{eq:f1}
     F_{\rm hz}(E_{j}) & \propto  & \int_{E_{j,\rm low}}^{E_{j,\rm high}} {E_{j}^\prime}^{-\Gamma_{\rm hz}} {\rm d} E_{j}^\prime \;,
    \label{eq:f2}
\end{eqnarray}
where $E_{j,\rm low}$ and $E_{j,\rm high}$ are the lowest and highest energy of the specific band of interest, respectively. 

Recently, \cite{Ingram2019} showed that there could be some misclassified photons in the energy band < 0.7 keV that should belong to higher energy bands due to the fact that {\it XMM-Newton}'s response matrix is non-diagonal. Taking into account this effect as well as the absorption column found in GX339--4 could change the flux normalization in a particular soft energy band. In this work, these effects are included indirectly by employing the reflected response fraction obtained from the spectral fits by \cite{Demarco2017} to produce the PSD, which will be described later in Section 4. Therefore, the PSD results here should tie to the mean spectrum fits of the data that have already folded through the instrument responses and that should incorporate the flux contamination induced by the effects of the response matrix as well.

\subsection{Mass accretion rate fluctuations and PSD}

When fluctuations in mass accretion rate, $\dot{m}$, propagate through the accretion disc, they induce variations in the seed photon emission. The disc seed photons are then Compton up-scattered inside the hot corona (or hot flow) producing the Comptonized photons whose variability is directly proportional to that of the mass accretion rate fluctuations. The observed continuum flux in the energy band $E_{j}$ then consists of two components associated with the emission from soft and hard zones, which can be described by
\begin{equation}
     x_{\rm 0}(t,E_{j}) = x_{\rm hz}(t,E_{j}) + x_{\rm sz}(t,E_{j})\;,
    \label{eq:lc}
\end{equation}
where 
\begin{eqnarray}
    x_{\rm hz}(t,E_{j}) & \propto & F_{\rm hz}(E_{j})\dot{m}(t) \;, \\
     x_{\rm sz}(t,E_{j}) & \propto & F_{\rm sz}(E_{j})\dot{m}(t+t_{\rm p})*h(t) \;.
    \label{eq:lc2}
\end{eqnarray}
$\dot{m}(t)$ is the variations of mass accretion rate fluctuations. $h(t)$ is the filter function that screens out the high-frequency variability of signals in the soft hot-flow zone so that the harder, inner zone produces the hard continuum spectra varying more at higher frequencies, which is the traditional property observed in PSD of AGNs \citep{Martin2012} and X-ray binaries \citep{Cui1997, Remillard2006}. Also, $h(t)$ produces the different X-ray variability in the two hot-flow components, so that the oscillatory interference features in the PSD can be produced. There is no one unique framework to explain distinct emission regions and the peaks seen in the PSD. This interference framework was previously proposed by \cite{Veledina2016, Veledina2018}, and was suggested by \cite{Mahmoud2018} that it can encompass other current models \citep[e.g.,][]{Ingram2012} within the parameter space \citep[see][for further discussion]{Mahmoud2019}. The variability produced from the soft hot-flow zone is modelled by a convolution ($*$ sign) of the filter function with the mass accretion rate. The parameter $t_{\rm p}$ is a characteristic propagation time from the $r_{\rm trc}$ to $r_{\rm sh}$ where the hard Comptonization operates. 

The PSD is estimated by the modulus squared of the discrete Fourier transform of the light curve \citep[e.g.,][]{Nowak1999,Emmanoulopoulos2013} which can be written as
\begin{equation}  
    P_{0}(f,E_{j}) \propto |X_{0}(f,E_{j})|^{2} /  \bar{x_{0}}^2  \;,
    \label{eq:psd}
\end{equation}  
where upper case letters represent the quantities in the frequency domain corresponding to those in the time domain written in lower case letters. According to eqs.~\ref{eq:lc}--\ref{eq:psd}, the PSD under this framework can be calculated via \citep[e.g.,][]{Veledina2016} 
\begin{equation}  
    P_{0}(f,E_{j}) = \frac{1+\epsilon^{2} H^{2}(f) + 2 \epsilon H(f)\cos(2\pi f t_{\rm p})   }{(1+\epsilon)^{2}}{|\dot{M}}(f)|^{2} \;,
    \label{eq:psd2}
\end{equation}  
where it is normalized in terms of the squared fractional rms \citep{Miyamoto1991} and $h(t)$ is a zero-lag function, so that $H(f)$ is real as this is instrumental in the derivation of eq.~\ref{eq:psd2}. $\epsilon = N_{\rm F}(F_{\rm sz}/F_{\rm hz})$ is the relative flux ratio of the X-ray continuum emitted from two different hot-flow zones. While $F_{\rm sz}/F_{\rm hz}$ depends on $\Gamma_{\rm sz}$, $\Gamma_{\rm hz}$, and the energy band being considered (see eqs.~1--2), we assume the normalization factor $N_{F}$ is the same for all energy bands in the same observation. The $N_{F}$ in the same energy band, however, could be different among different observations. $H(f)$ is the Fourier form of the filter function which can be expressed in the form \citep{Veledina2016}
\begin{equation}  
    H(f) = \frac{1}{(f/f_{\rm filt})^{4}+1} \;
    \label{eq:ffilt}
\end{equation}  
so that the transmitted signals above the frequency $f_{\rm filt}$ are damped. 

The variability power of the mass accretion rate significantly decays at the viscous frequency \citep{Ingram2013}  
\begin{equation}
    f_{\rm visc}(r_{n}) = r_{n}^{-3/2}(H/R)^{2}\alpha/2\pi \;,
    \label{eq:fvis}
\end{equation}
where $H/R$ is the disc scale-height ratio and $\alpha$ is the viscosity parameter. The radius $r_{\rm n}$ is in units of $r_{\rm g}$ so the viscous frequency $f_{\rm visc}(r_{n})$ is in units of $1/t_{\rm g}$, where $t_{\rm g} = GM/c^{3}$. One can convert gravitational units to physical units when the black hole mass is known. For a black hole mass of $10M_{\odot}$, $1r_{\rm g} = 1.48\times10^{4}$~m and $1t_{\rm g} = 4.9\times10^{-5}$~s. 

To describe the broken power-law shape usually seen in the PSD, the form of $|{\dot{M}}|^{2}(f)$ is assumed to be
\begin{equation}
    |{\dot{M}}(f)|^{2} = \frac{1}{1+(f/f_{\rm break})^{\beta}}\;,
    \label{eq:mdot_f}
\end{equation}
where $f_{\rm break}$ is the break frequency and $\beta$ is an arbitrary index. We fix $\beta=1.5$ when investigate other model parameters, but allow it to be free when we fit the data. ${\dot{M}}(f)$ represented in eq.~\ref{eq:mdot_f} is then the frequency domain form of the driving signal that is used to activate the X-ray continuum emission and variability from the hot-flows.

There are two important parameters in our model giving the radial positions of the truncation radius, $r_{\rm trc}$, and the transition radius where the hot flows change from the spectrally soft to spectrally hard zone, $r_{\rm sh}$. We then relate the break and the filter frequency to the viscous timescales at the truncation and transition radius, respectively,
\begin{eqnarray}
f_{\rm break} & = &f_{\rm visc}(r_{\rm trc})\;, \\
f_{\rm filt} & = &f_{\rm visc}(r_{\rm sh})\;.
\label{eq:fbff}
\end{eqnarray}
The characteristic time for fluctuations that originate at a radius $r$ to propagate across a radial distance $\Delta r$ is $\Delta t =\Delta r /[r f_{\rm visc}(r)]$ \citep{Ingram2013}. Assuming the characteristic time for fluctuations to propagate through the soft zone depends on the viscous timescale at the truncation radius, the value of $t_{\rm p}$ in eqs.~\ref{eq:lc2} and \ref{eq:psd2} then can be estimated as
\begin{equation}
     t_{\rm p} = (r_{\rm trc} - r_{\rm sh})/(r_{\rm trc} f_{\rm break})\;.
    \label{eq:tp}
\end{equation} 

\begin{figure}
    \centerline{
        \includegraphics[width=0.45\textwidth]{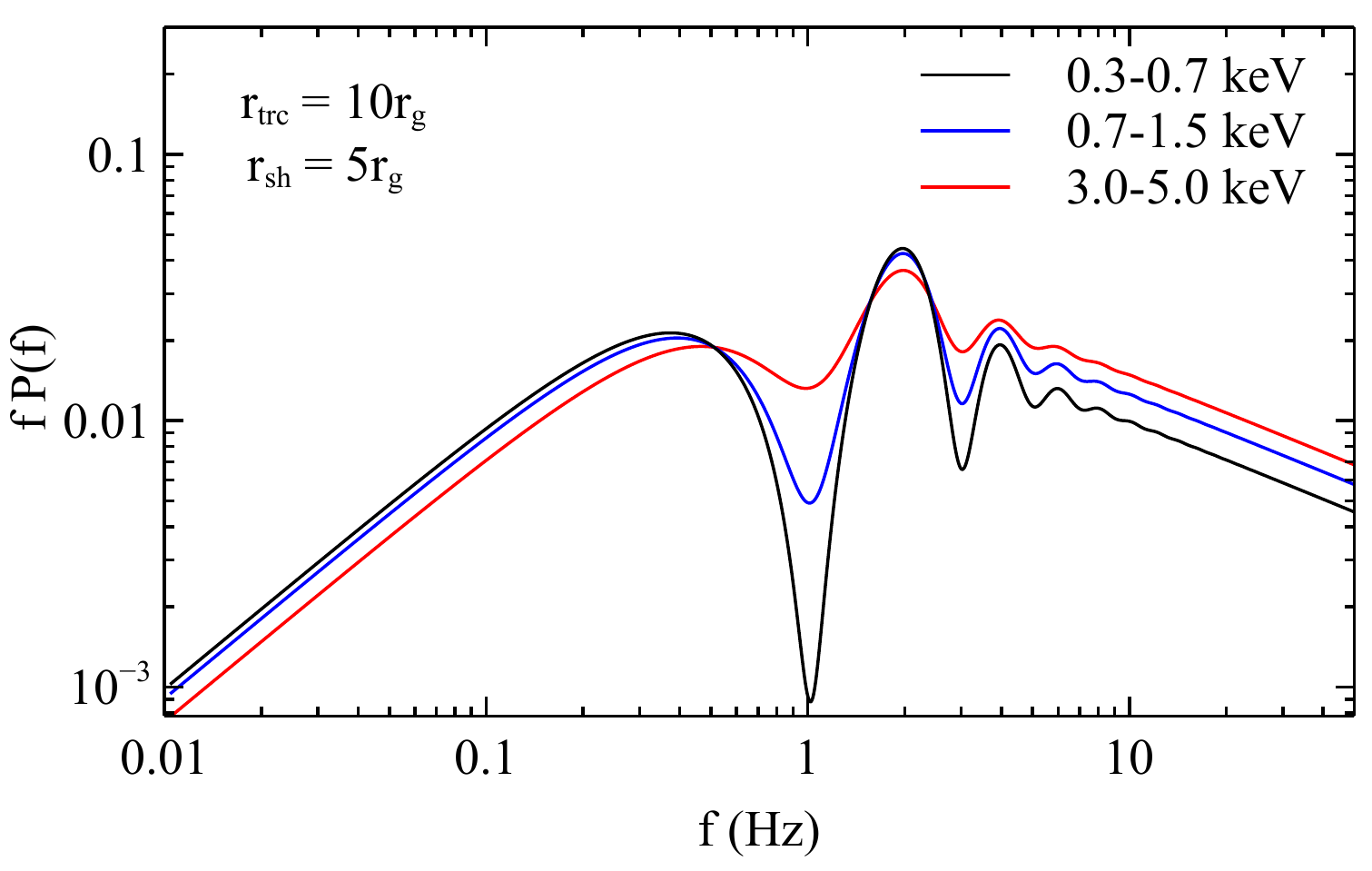}
    }
    \caption{The PSD in the 0.3--0.7~keV energy band (black line), 0.7--1.5~keV (blue line), and 3--5~ keV (red line) with $r_{\rm trc}=10r_{\rm g}$ and $r_{\rm sh}=5r_{\rm g}$. The soft and hard zones of the inner flow have the spectral photon indices $\Gamma_{\rm s}=2.5$ and $\Gamma_{\rm h}=2.0$, respectively. We fix $N_{\rm F}=0.5$ so that all energy bands are different due to the values of the intrinsic flux ratio $F_{\rm s}/F_{\rm h}$.  
    }
    \label{fig2}
\end{figure}

Fig.~\ref{fig2} shows examples of the model PSD for different energy bands when the photon indices of the continuum emitted from soft and hard zones are $\Gamma_{\rm sz} = 2.5$ and $\Gamma_{\rm hz} = 2.0$, respectively. We assume the disc is truncated at $r_{\rm trc}=10r_{\rm g}$ and the transition radius is $r_{\rm sh}=5r_{\rm g}$. The disc parameter $(H/R)^{2}\alpha$ is set to be 0.01. The model can produce larger high-frequency power in higher energy bands, which is the traditional PSD properties seen in both AGN and X-ray binaries. Furthermore the dips and humps are naturally produced by the model. We fix $N_{\rm F}=0.5$ so the parameter $\epsilon = N_{\rm F}(F_{\rm sz}/F_{\rm hz})$ that regulates the importance of the filter level depends only on $F_{\rm sz}/F_{\rm hz}$, which is different in each energy band. Generally, the dips and humps could also depend on $r_{\rm trc}$ and $r_{\rm sh}$ that determine $f_{\rm break}$, $f_{\rm filt}$, and the characteristic propagation time $t_{\rm p}$.

\begin{figure}
    \centerline{
        \includegraphics[width=0.45\textwidth]{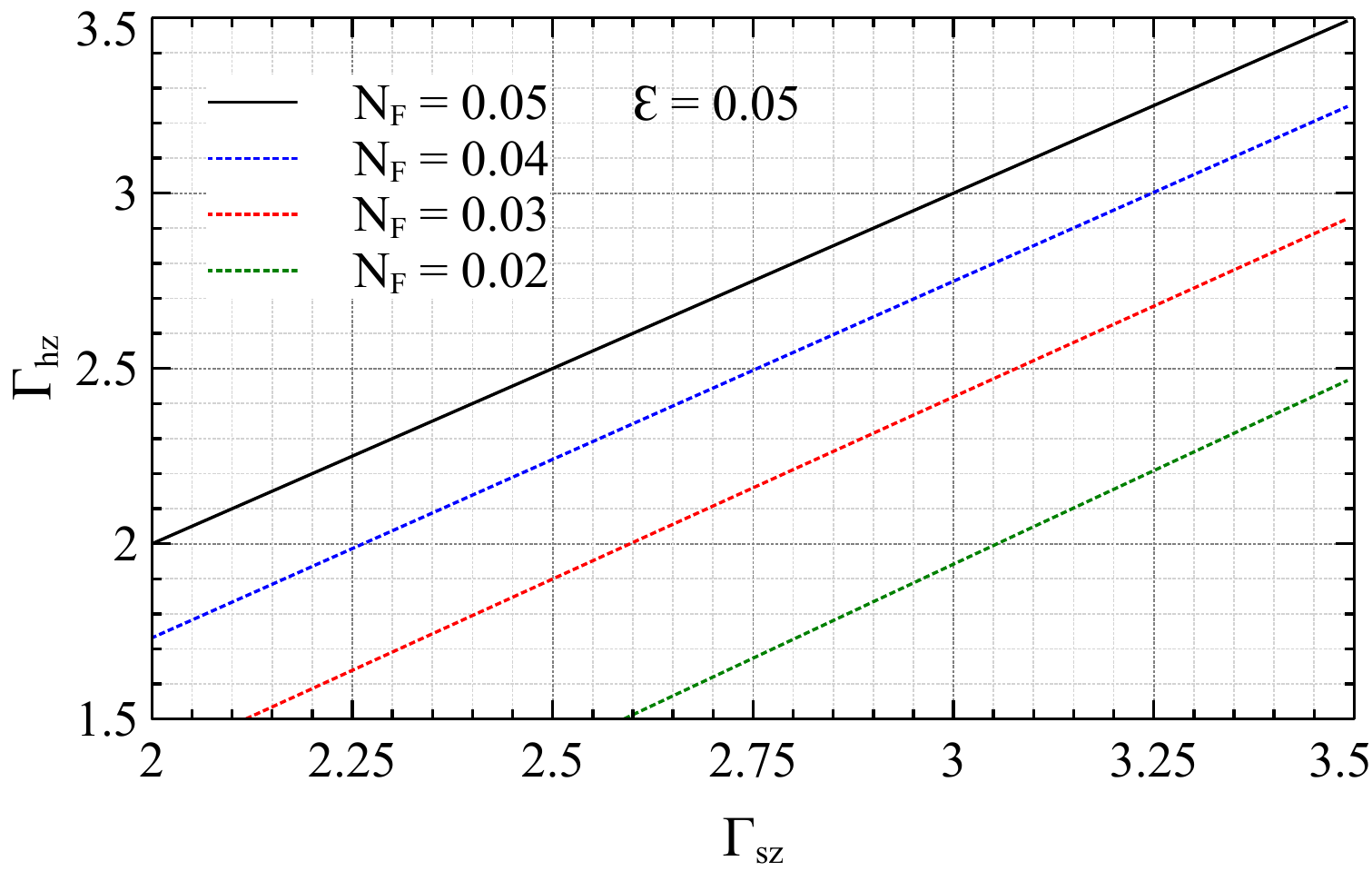}
    }
    \vspace{0.2cm}
    \centerline{
        \includegraphics[width=0.45\textwidth]{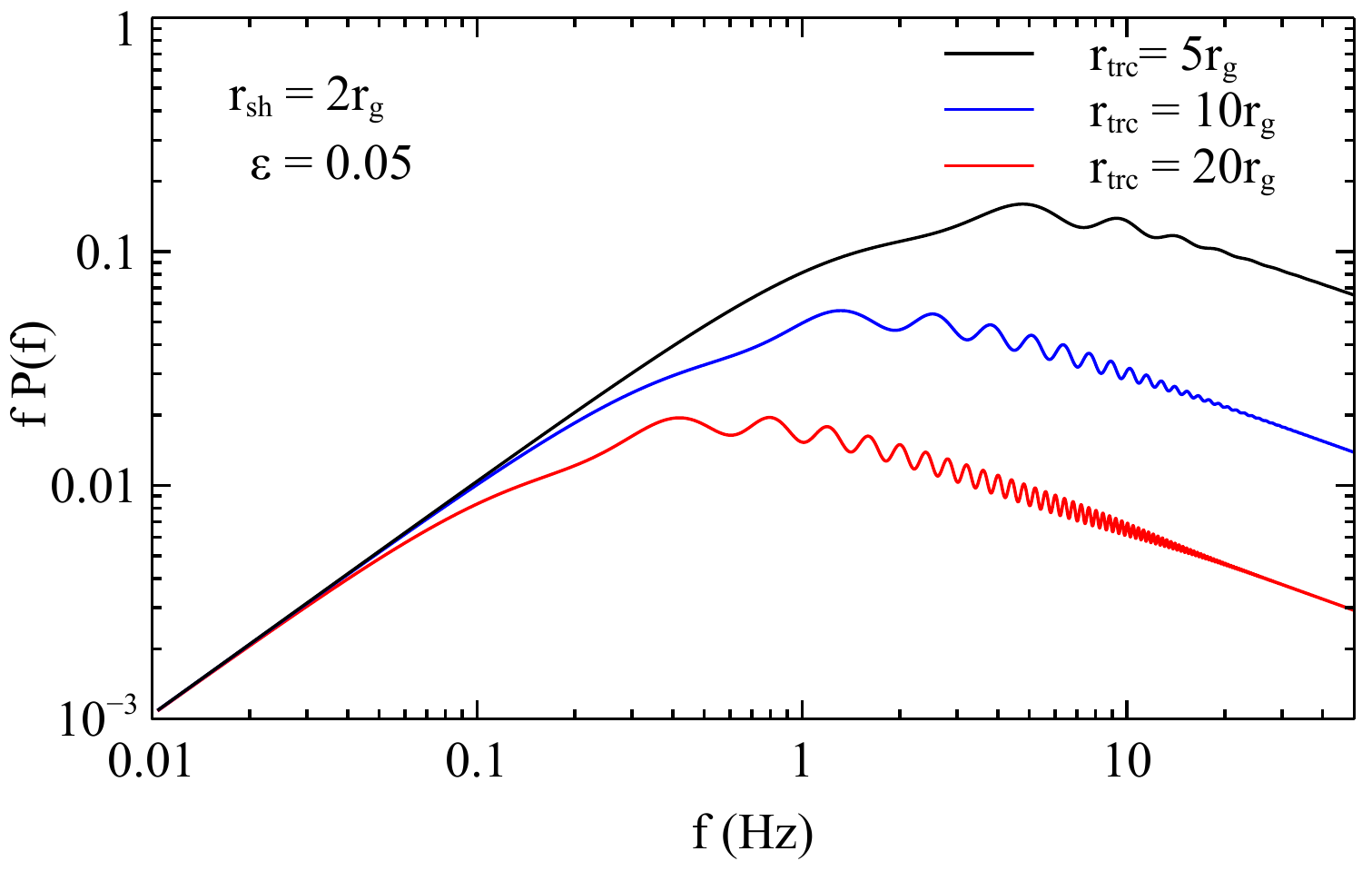}
    }
    \vspace{0.2cm}
     \centerline{
        \includegraphics[width=0.45\textwidth]{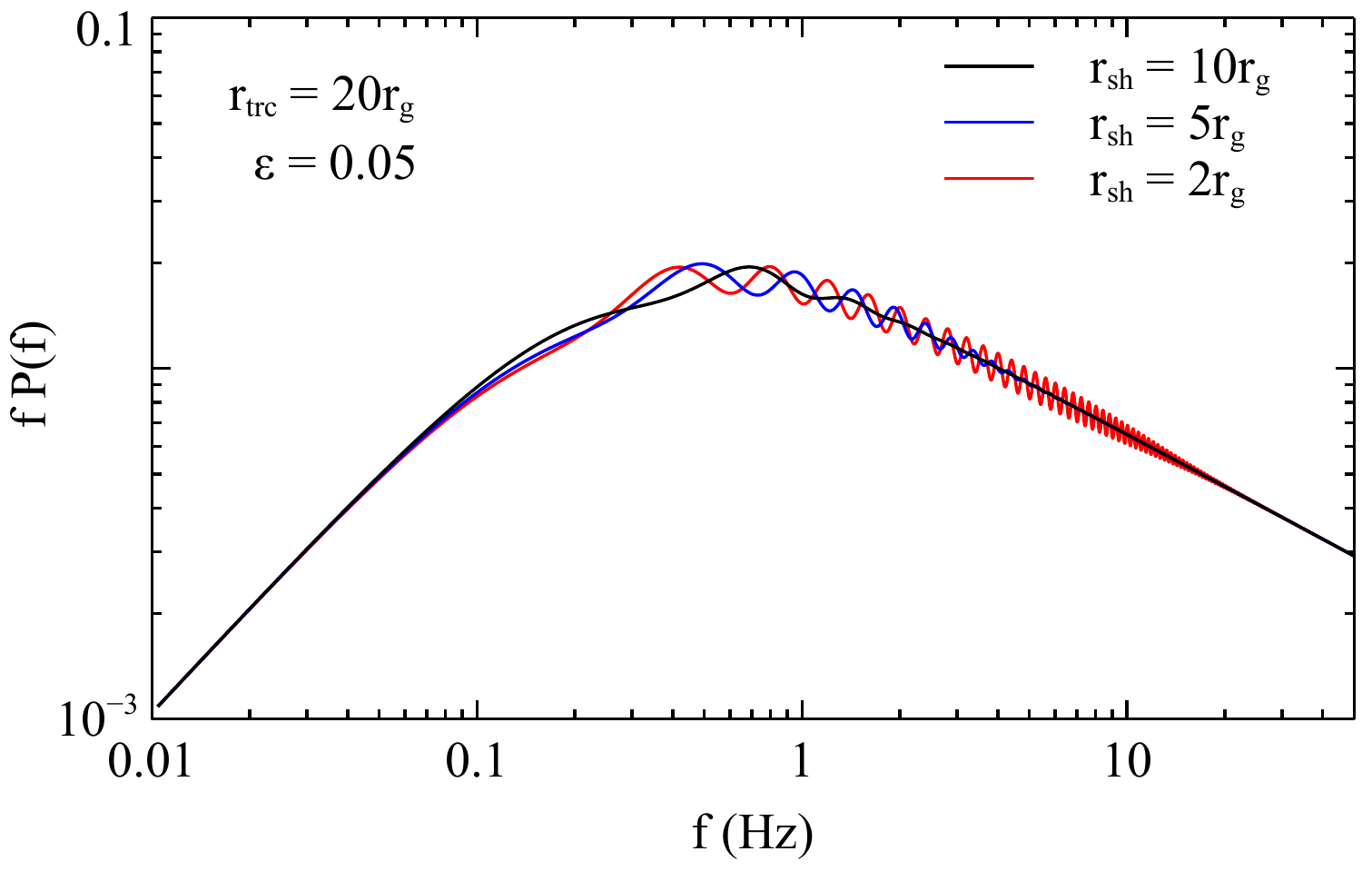}
    }
    \caption{\emph{Top panel}: $\Gamma_{\rm sz}$ vs. $\Gamma_{\rm hz}$ plots that give $\epsilon=0.05$ for different $N_{\rm F}$ values for a 0.3--0.7 keV band of interest. \emph{Middle panel}: The corresponding PSD in 0.3--0.7~keV band varying with $r_{\rm trc}$. \emph{Bottom panel}: The corresponding PSD in the same band varying with $r_{\rm sh}$.}
    \label{fig3}
\end{figure}

How the geometry of the inner flows affects the PSD in 0.3--0.7 keV band when $\epsilon=0.05$ and $(H/R)^{2}\alpha=0.01$ is shown in Fig.~\ref{fig3}. We note that when $\epsilon$ is fixed, there can be different pairs of $\Gamma_{\rm sz}$ and $\Gamma_{\rm hz}$ that produce the same PSD profile for each value of $N_{\rm F}$. To illustrate this, Fig.~\ref{fig3} (top panel) shows the $\Gamma_{\rm sz}$ v.s. $\Gamma_{\rm hz}$ plots that provide $\epsilon=0.05$ for different $N_{\rm F}$. In principle, we expect $\Gamma_{\rm sz} > \Gamma_{\rm hz}$, so the black solid line marks the upper limit of $N_{\rm F}$ of which above this value the model returns $\Gamma_{\rm sz} < \Gamma_{\rm hz}$ in this energy band, which is unrealistic and can be neglected. Fig.~\ref{fig3} (middle panel) shows that stronger high-frequency power is obtained with decreasing $r_{\rm trc}$. Since we fix $r_{\rm sh} = 2r_{\rm g}$, larger $r_{\rm trc}$ results in longer propagation-time delay, $t_{\rm p}$, through the inner soft to the inner hard hot-flow zones. Increasing time $t_{\rm p}$ results in more humps being imprinted on the PSD due to the effect of the cosine term in eq.~\ref{eq:psd2}. This is also true when we fix $r_{\rm trc}=20r_{\rm g}$, but vary $r_{\rm sh}$ and hence also $t_{\rm p}$ (Fig.~\ref{fig3}, bottom panel). Moreover, the smaller $r_{\rm sh}$ is, the higher frequency the oscillations is significantly filtered. This is because the $f_{\rm filt}$ is larger for smaller $r_{\rm sh}$. Decreasing $r_{\rm sh}$ means the inner-soft zone could produce X-rays on shorter characteristic timescales to interference with X-rays from the hard hot-flows, leading to the oscillation, interference structures on the PSD profiles towards higher frequencies.   

\subsection{Simulated PSD including reverberation}

Since the radial size of the disc is relatively large comparing to that of the inner hot flows, the source illuminating the disc can be approximated as a central illuminating source \citep[e.g.][]{Gardner2014, Mahmoud2019}. Therefore, the time delay of photons travelling from the central source to be reprocessed by the disc at radius $r$ is given by \citep{Welsh1991} 
\begin{equation}
    \tau = \frac{r}{c}\bigg{(}1- \sin{i} \cos{\phi} \bigg{)}  \;,
    \label{eq:impuse_res}
\end{equation}
where $i$ is the inclination angle measured from the observer’s line of sight to the disc axis and $\phi$ is the azimuthal angle between each specific point on the disc and the projection of the line of sight onto the disc. We note that eq.~\ref{eq:impuse_res} does not include general relativistic effects but is still acceptable since we do not specify the exact geometry of the flows. The impulse response function due to reverberation, $\psi(\tau)$, can be produced by summing the photon counts over a given range of radii $r_{\rm trc} \leq r \leq 400r_{g}$ and all azimuthal angles $0 \leq \phi \leq 2\pi$ as a function of the time delay using eq.~\ref{eq:impuse_res}. At radii smaller than $r_{\rm trc}$, the disc is replaced by the hot flow and X-ray reverberation does not take place.

\begin{figure}
    \centerline{
        \includegraphics[width=0.45\textwidth]{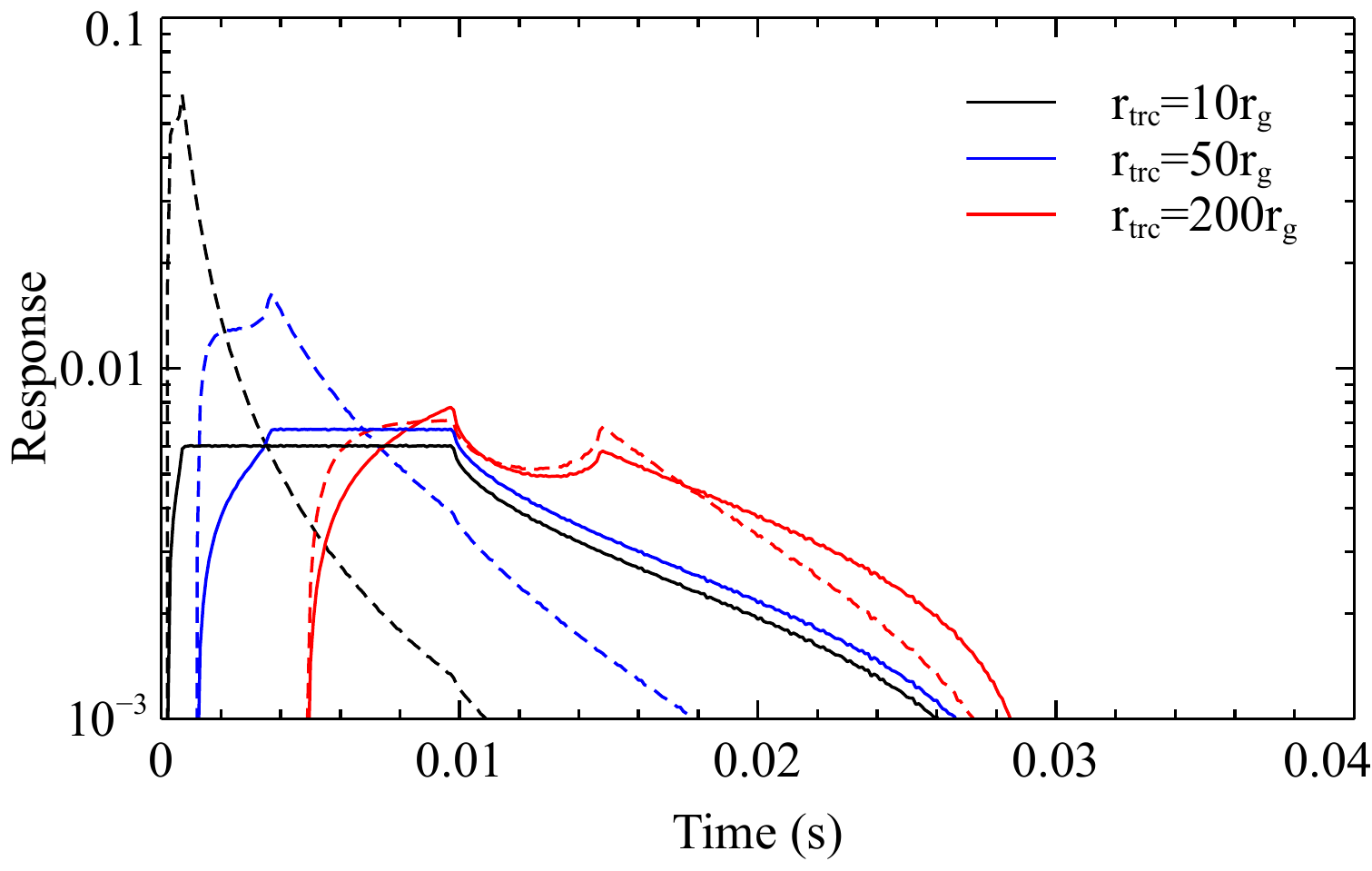}
    }
    \caption{Simulated impulse disc responses in 0.3--0.7~keV band varying with the disc truncation radius. The solid and dotted lines represent the cases when the emissivity index $\gamma=0$ and 1.5, respectively. The inclination angle is fixed at $i=30^{\circ}$ and $M_{\rm BH}=10M_{\odot}$.}
    \label{p4}
\end{figure}

We introduce the parameter $\gamma$ to describe the emissivity of the accretion disc so that the number of emitted photons from each annulus of area $2 \pi r {\rm d} r$ is proportional to $r^{-\gamma}$. Higher $\gamma$ means the X-ray reflection is more concentrated at the inner part of the disc. Meanwhile, $\gamma=0$ means that the same number of photons is produced in each area of $2 \pi r {\rm d} r$ and that emissivity, in terms of flux per unit area on the disc, is proportional to $r^{-1}$. Examples of the disc-response functions are presented in Fig.~\ref{p4}. Here we fix $M_{\rm BH}=10M_{\odot}$ and $i=30^{\circ}$ while allow the $r_{\rm trc}$ to be a free parameter. The solid and dotted lines represent the case of $\gamma=0$ and $1.5$, respectively. 

The impulse reverberation response also acts as a filter on the driving signal that affects the shape of the observed PSD \citep{Papadakis2016, Chainakun2019a}. According to the convolution theorem, the Fourier transform of an observed light curve including reverberation effects can be written as  \citep{Uttley2014, Papadakis2016} 
\begin{equation}
    X(f,E_{j}) = X_{0}(f,E_{j})\Psi(f,E_{j}) \;,
\end{equation}
where $X_{0}(f)$ and $\Psi(f)$ are the Fourier transform of the driving signal and the impulse response, respectively. The observed PSD in the energy band $E_{j}$ is then given by 
\begin{equation}
   P_{\rm obs}(f, E_{j}) = \frac{|X(f,E_{j})|^{2}}{\big{(}1+R(E_{j})\big{)}^{2}} = \frac{|X_{0}(f,E_{j})|^{2}|\Psi(f,E_{j})|^{2}}{\big{(}1+R(E_{j})\big{)}^{2}}\; ,
 \end{equation}
where $R(E_{j})$ is the reflected response fraction defined as the (reflection flux)/(continuum flux) measured in the specific energy band $E_{j}$, which can vary between observations. 

\begin{figure}
    \centerline{
        \includegraphics[width=0.45\textwidth]{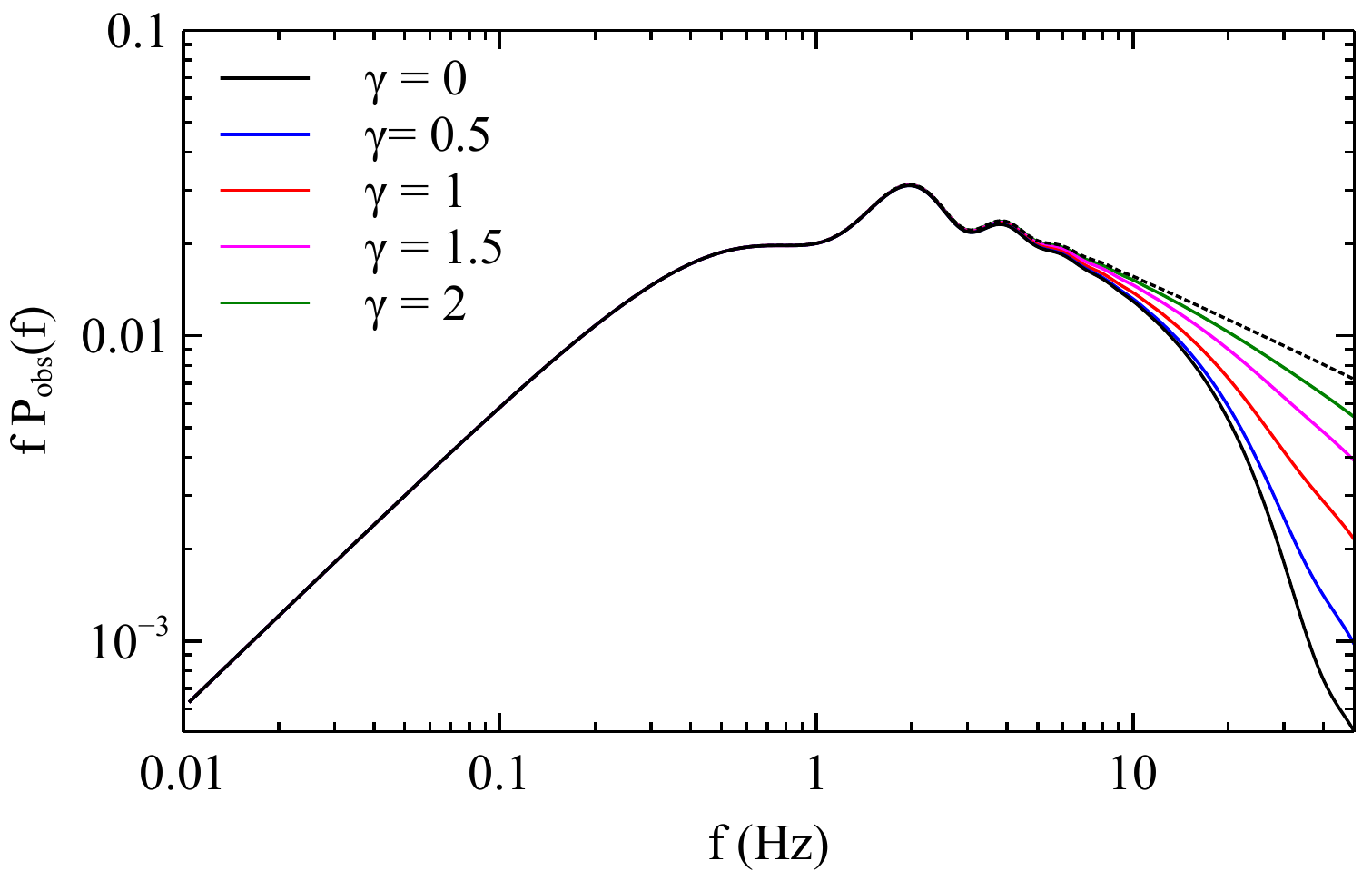}
    }
    \vspace{0.2cm}
    \centerline{
        \includegraphics[width=0.45\textwidth]{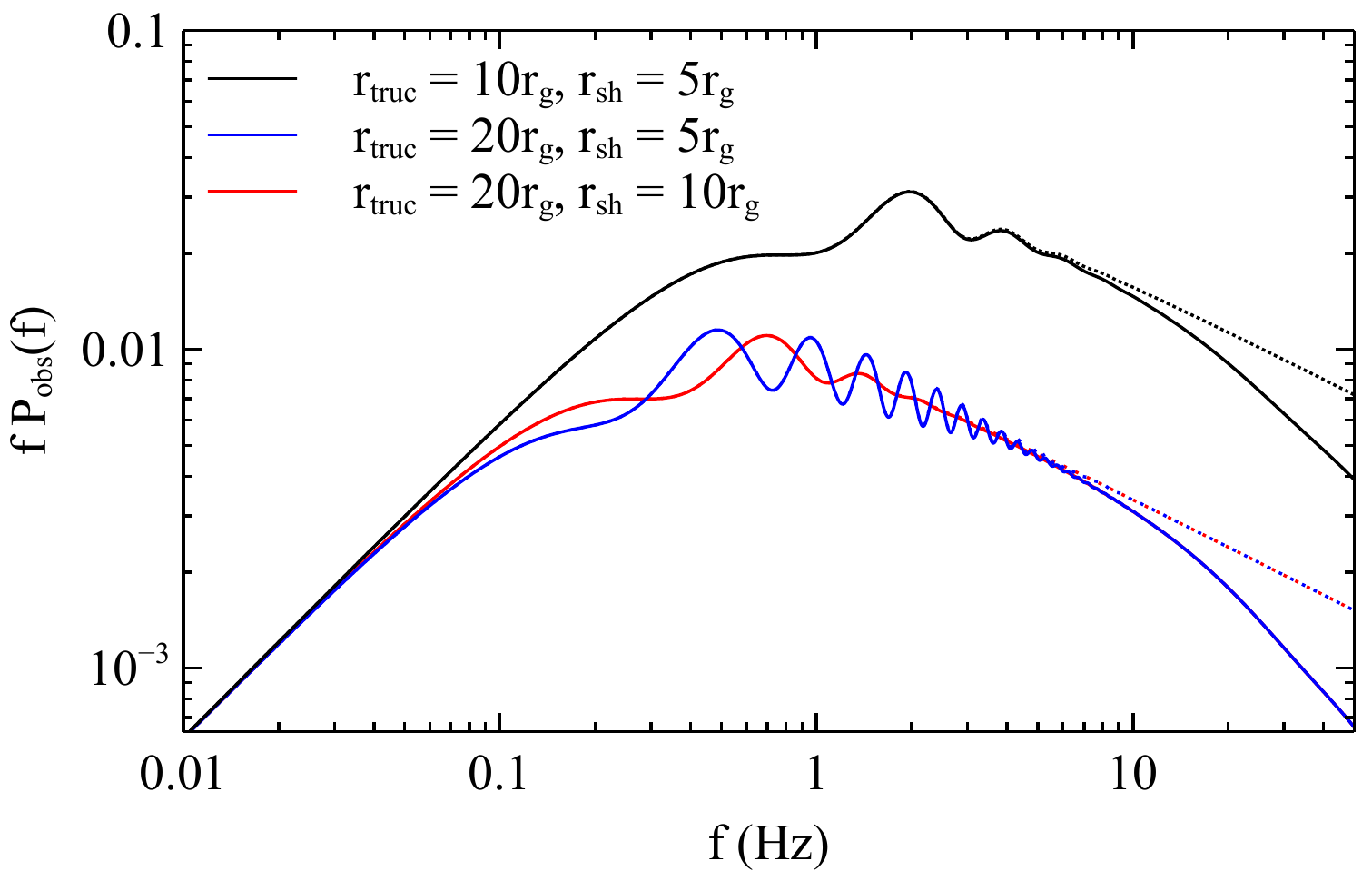}
    }
    \caption{\emph{Top panel}: The total PSD in 0.3--0.7~keV band varying with $\gamma$ when $i=30^{o}$, $r_{\rm trc}=10r_{\rm g}$, $r_{\rm sh}=5r_{\rm g}$, $\epsilon=0.1$ and $(H/R)^{2}\alpha=0.01$. \emph{Bottom panel}: The total PSD in 0.3--0.7~keV band varying the geometry of the system with $\gamma=1.5$. All dotted lines represents the corresponding cases with reverberation excluded. The reduction of power at high frequencies is produced by reverberation.}
    \label{p5}
\end{figure}

The 0.3--0.7 keV PSD predicted by the model that includes both effects of disc fluctuations and reverberation are presented in Fig.~\ref{p5} (solid lines). We fix $\epsilon=0.1$ and $(H/R)^{2}\alpha=0.01$, but vary $\gamma$, $r_{\rm trc}$ and $r_{\rm sh}$. For comparison, the corresponding PSD with reverberation excluded is also shown as the dotted lines. It is clear that the reverberation acts as a filter that reduces the high-frequency power in the observed PSD, which is independent of the input signals \citep[see, also, discussion in][]{Papadakis2016,Chainakun2019a}. Interestingly, if the emissivity $\gamma$ is too high (i.e., the source illumination is too centrally concentrated), most photons would reflect from inner disc so the reverberation delays would be quite short and, consequently, relatively small effects of reverberation are seen in the PSD profiles towards low temporal frequencies (i.e., long timescales). We note that the reverberation features are imprinted on the PSD at high temporal frequencies associated with the short timescales of the disc reflection. Given $M_{\rm BH}=10M_{\odot}$, the significant drop of power due to reverberation at frequencies around a few tens of Hz should be the beginning of the main dip of the oscillatory structures in the PSD produced by reverberation such like in case of AGN \citep{Papadakis2016, Chainakun2019a}. Furthermore, the result in Fig.~\ref{p5} (bottom panel) suggest that the overall features of the PSD are quite different for different combinations of key model parameters such as $r_{\rm trc}$, $r_{\rm sh}$, $\epsilon$ and $\gamma$.

\section{Fitting results}

Our model consists of nine parameters in total: truncation radius, $r_{\rm trc}$, soft-to-hard transition radius, $r_{\rm sh}$, disc parameter, $(H/R)^{2}\alpha$, normalization flux, $N_{\rm F}$, photon index of the hard zone, $\Gamma_{\rm hz}$, photon index of the soft zone, $\Gamma_{\rm sz}$, low-frequency break index, $\beta$, reflected response fraction, $R(E_{j})$, and disc emissivity index, $\gamma$. To minimize the free parameters, the $R(E_{j})$ is fixed to the values obtained from the spectral fits by \cite{Demarco2017}, which are approximately 4.0 and 1.0 in 0.3--0.7 keV and 0.7--1.5 keV bands, respectively. The reflected response fractions employed here describe the effects of contamination flux between cross-components in the reverberation-dominated and continuum-dominated energy bands. In principle, they should also be able to describe the effect of misclassifying photons due to the non-diagonal response matrix of the instrument that plays a role in changing contamination flux as well. The effects of changing the photon index on the PSD are similar to changing $N_{\rm F}$, which is a free parameter. Since the photon index of the X-ray continuum in these GX~339--4 observations was previously reported to vary between 1.5--2 \citep[e.g.,][]{Demarco2017}, we fix $\Gamma_{\rm hz} = 1.5$ and $\Gamma_{\rm sz} = 2.0$. This helps avoid the degeneracies in the model and also avoid predicting a spectrum that is wildly different from the observed spectrum. The parameter $N_{\rm F}$ is fixed between two energy bands, but allowed to vary across the observations. 

\begin{figure*}
\centerline{
\includegraphics*[width=0.55\textwidth]{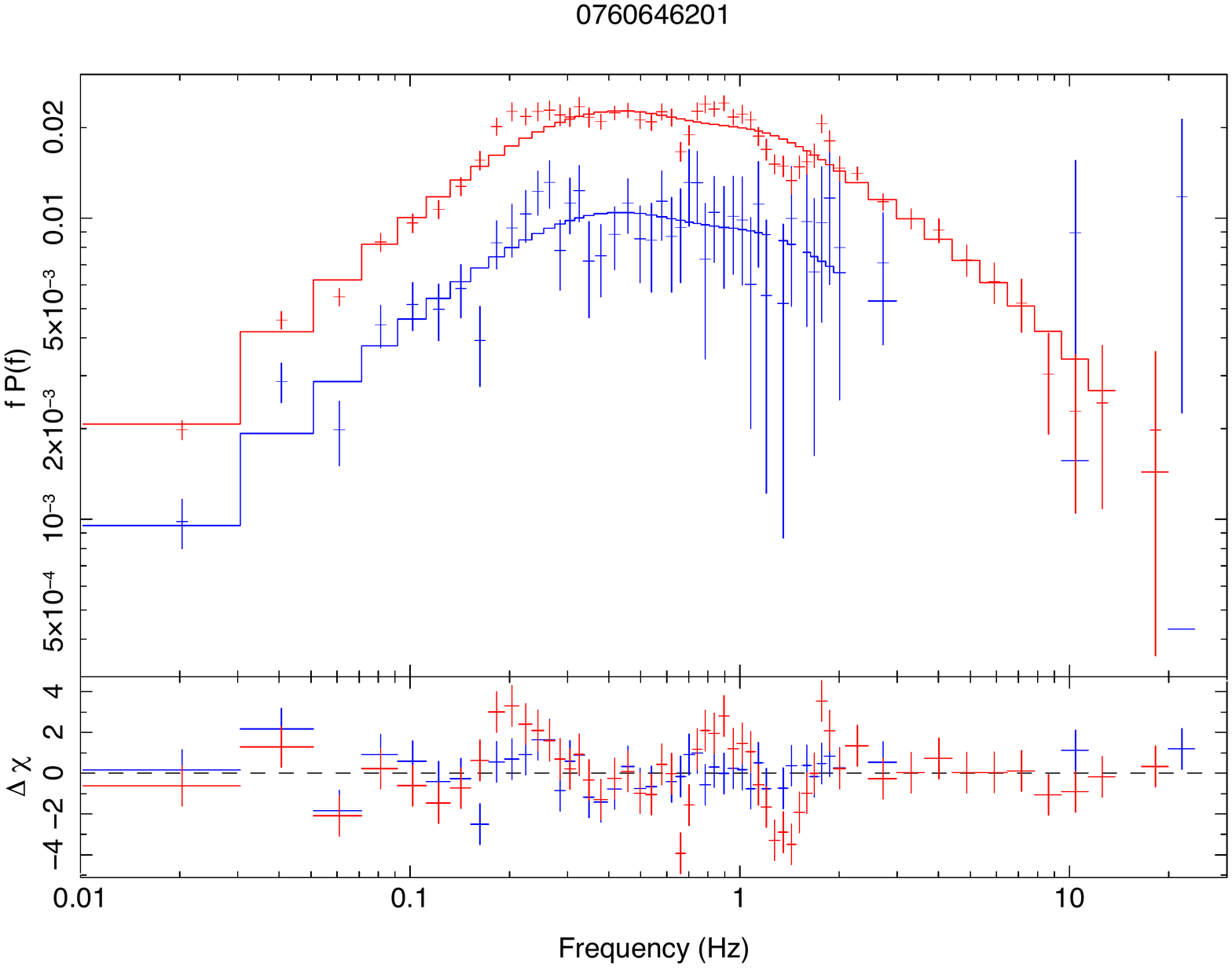}
\hspace{-1.0cm}
\includegraphics*[width=0.55\textwidth]{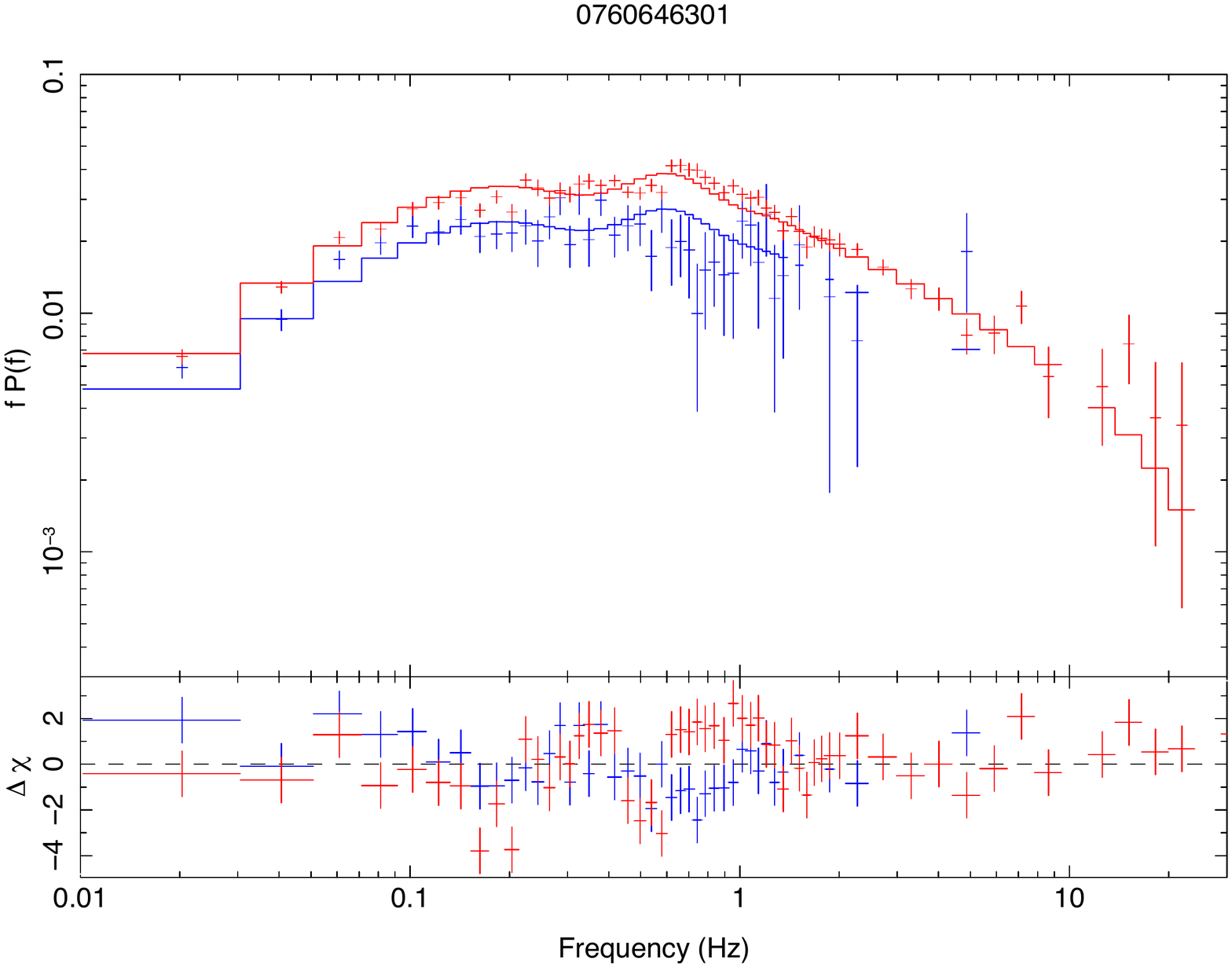}
\vspace{-0.5cm}
}
\centerline{
\includegraphics*[width=0.55\textwidth]{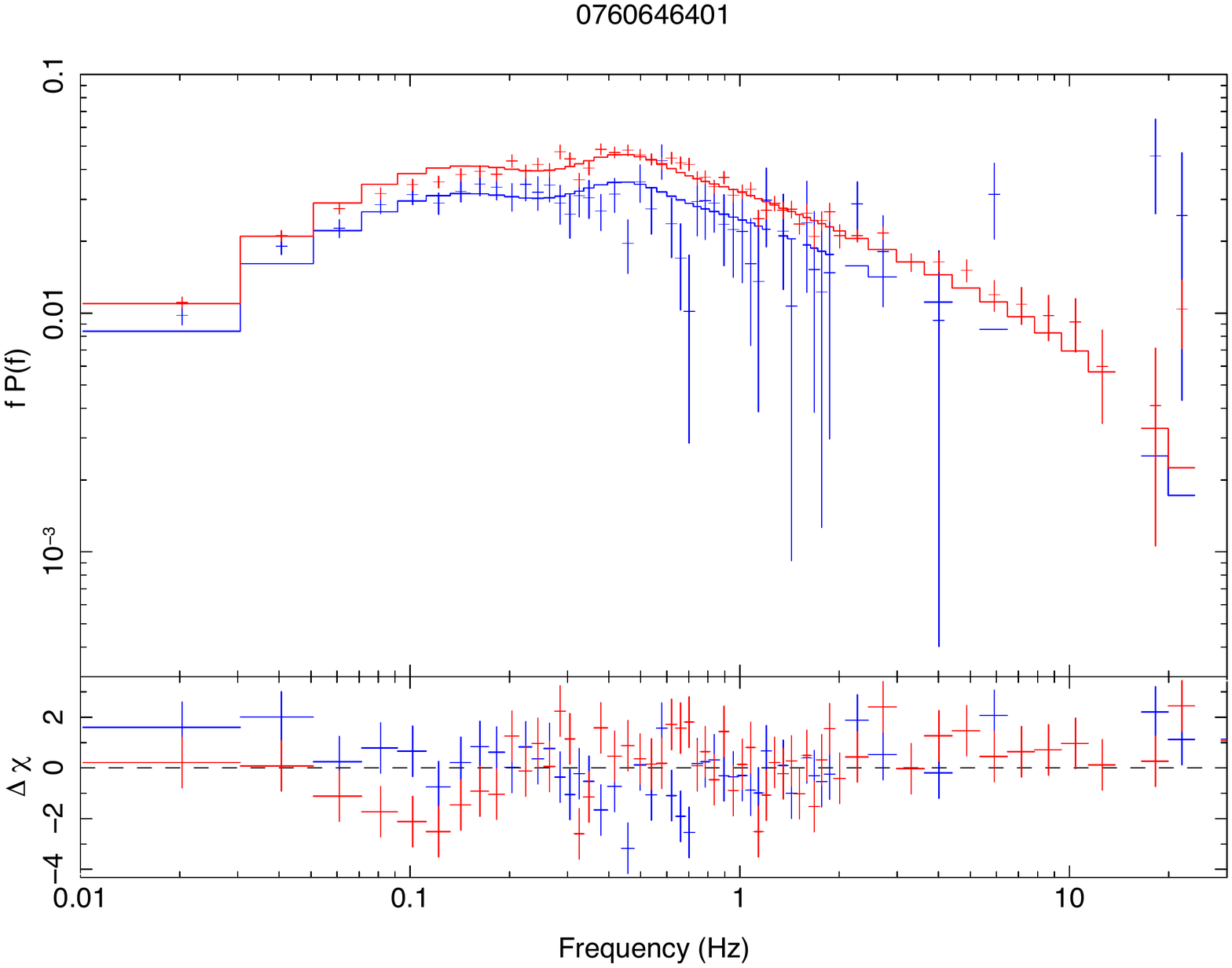}
\hspace{-1.0cm}
\includegraphics*[width=0.55\textwidth]{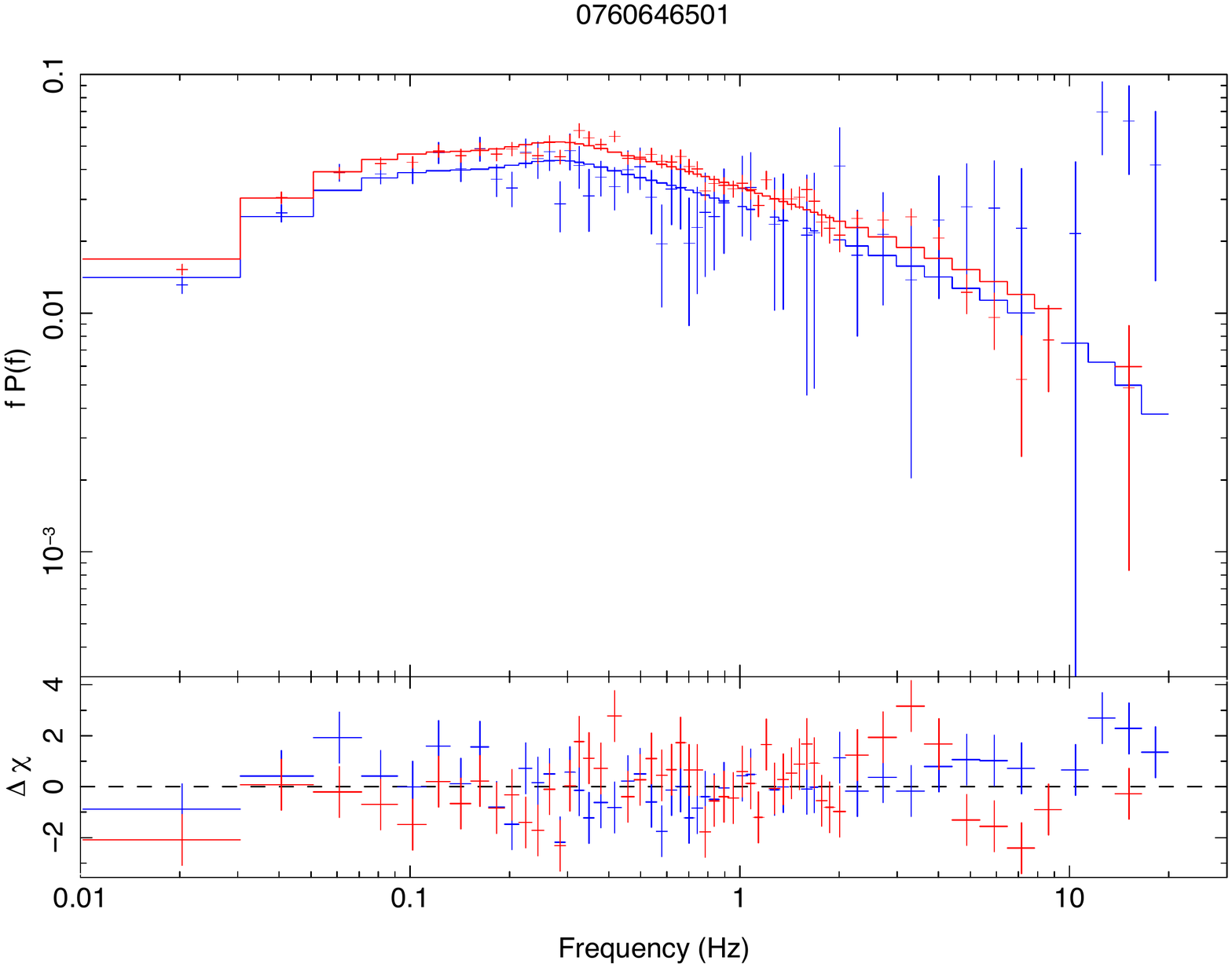}
\vspace{-0.5cm}
}
\centerline{
\includegraphics*[width=0.55\textwidth]{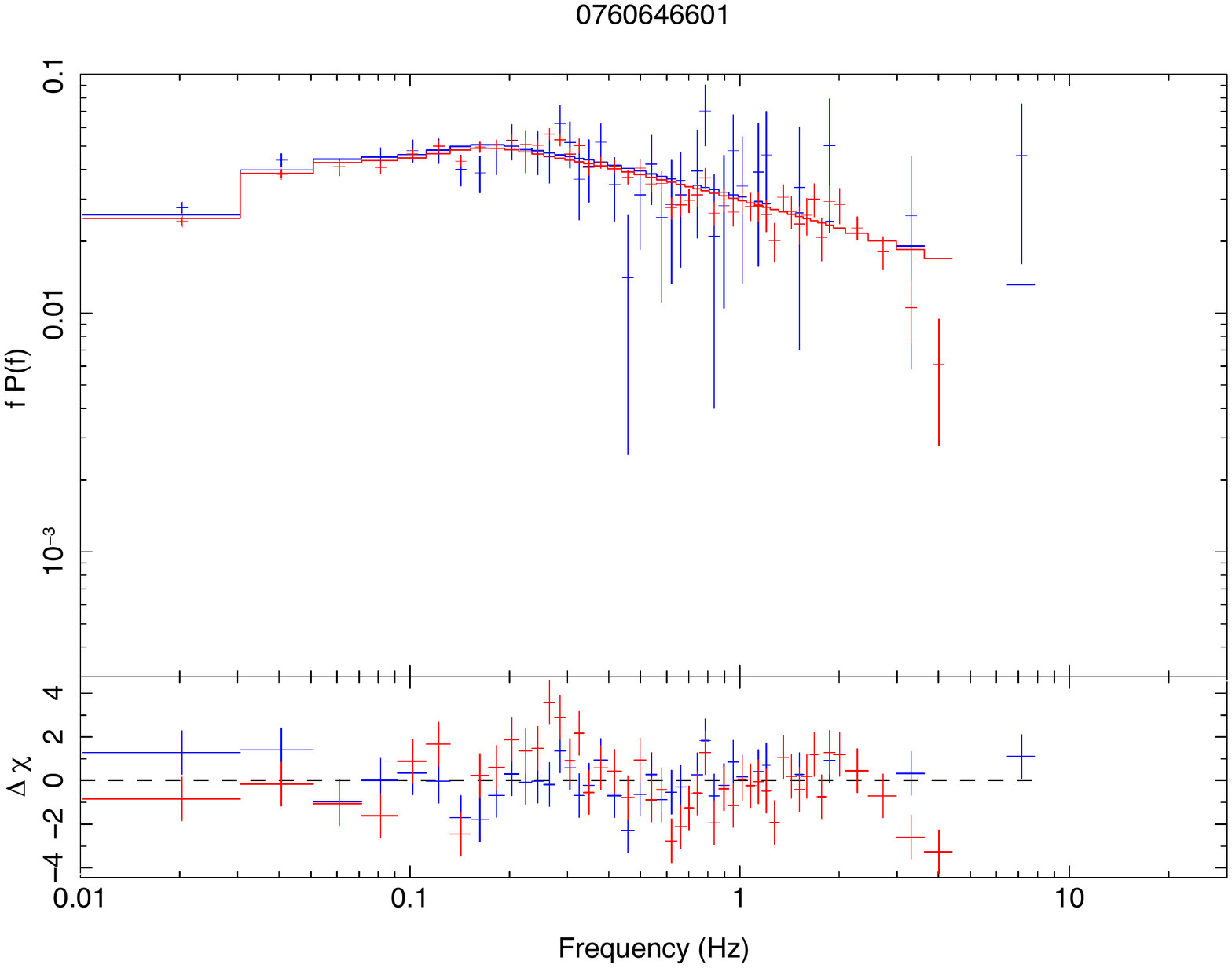}
\hspace{-1.0cm}
\includegraphics*[width=0.55\textwidth]{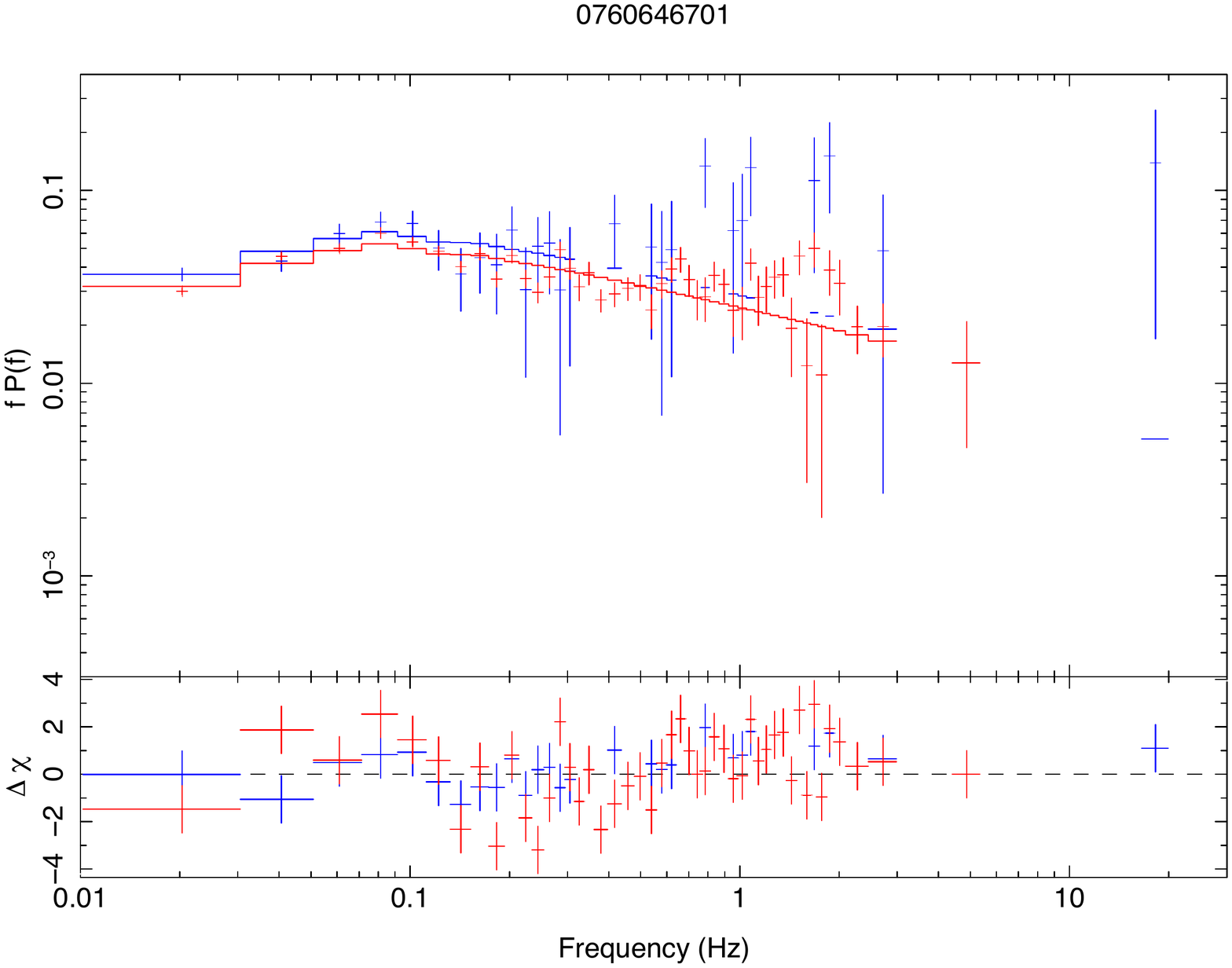}
}
\caption{Data, model and residuals from fitting the PSD model to 0.3--0.7 keV (blue) and 0.7--1.5 keV (red) PSD data with $(H/R)^{2}\alpha$ fixed at 0.005 and $\gamma$ fixed at 0 as obtained by the global fits. The PSD displayed here are Poisson noise-subtracted and normalized in terms of squared fractional rms \citep{Miyamoto1991}. Data points consistent with zero (or negative) power have been excluded. The best-fit model parameters for each observation are listed in Table~\ref{tab_fit_para}. }
\label{fig_best_fit}
\end{figure*}

According to \cite{Veledina2016}, the disc and viscous parameters of GX~339--4, via modelling the PSD in 2--15 keV, were found to be $H/R > 0.23$ and $0.01 < \alpha < 0.5$. We consider the cases when $0.0005 \lesssim (H/R)^{2}\alpha \lesssim 0.1$ which is approximately between the lower limit and a reasonable upper limit $(H/R)=1$ and $\alpha=0.1$, but still in the range of what previously reported \citep{Veledina2016}. Firstly, the global grids of the model have been produced and the PSD data in two energy bands are simultaneously fitted in ISIS \citep{Houck2000}. We find that for the majority of fits the values of $(H/R)^{2}\alpha$ is 0.005 and $\gamma$ is 0. To improve the fit while limiting computation time, finer local grids of the model are produced independently for each observation, but this time both $(H/R)^{2}\alpha$ and $\gamma$ are fixed at the values constrained by the global grids, under the assumption that they do not vary much during these observations. The fitting is repeated with the finer, local grids for each observation with fixed $(H/R)^{2}\alpha = 0.005$ and $\gamma = 0$. The fitting results are presented in Fig.~\ref{fig_best_fit}. The corresponding best-fit parameters are listed in Table~\ref{tab_fit_para}. We found an increase in the truncation radius from $\sim 10r_{\rm g}$ to $\sim 55r_{\rm g}$ during O1--O6 as the source is decreasing in flux towards the end of the outburst. An increase of the soft--hard transition radius and a variation of $N_{\rm F}$ are also found during O1--O6. Meanwhile, $\beta$ seems to decrease towards the end of the outburst.

\begin{table*}
\begin{center}
 \caption{Fit parameters for GX~339--4 with $(H/R)^2 \alpha = 0.005$ and $\gamma = 0$.} 
 \label{tab_fit_para}
\begin{tabular}{lllllll}
\hline
Parameter & O1 & O2 & O3 & O4 & O5 & O6 \\
\hline
$r_\text{trc} (r_g)$ & $10^{+1.14}_{-0.89}$ & $15^{+1.16}_{-1.89}$ & $19^{+0.19}_{-0.20}$ & $26^{+0.16}_{-0.00}$ & $39^{+1.12}_{-1.34}$ & $55^{+2.38}_{-3.98}$ \\
$r_\text{sh} (r_g)$ & $7^{+0.64}_{-0.75}$ & $9^{+1.24}_{-1.93}$ & $12^{+0.47}_{-0.20}$ & $17^{+1.48}_{-0.50}$ & $27^{+0.13}_{-1.36}$ & $29^{+2.90}_{-3.20}$ \\
$(H/R)^2 \alpha$ & 0.005$^f$ & 0.005$^f$ & 0.005$^f$ & 0.005$^f$ & 0.005$^f$ & 0.005$^f$ \\
$\beta$ & $1.8^{+0.01}_{-0.19}$ & $1.7^{+0.01}_{-0.20}$ & $1.6^{+0.01}_{-0.20}$  & $1.5^{+0.00}_{-0.08}$ & $1.4^{+0.03}_{-0.10}$ & $1.4^{+0.02}_{-0.08}$ \\
$\gamma$ & 0$^f$  & 0$^f$ & 0$^f$ & 0$^f$ & 0$^f$ & 0$^f$ \\
$N_F$ & $0.05^{+0.01}_{-0.03}$ & $0.08^{+0.01}_{-0.07}$ & $0.07^{+0.01}_{-0.03}$ & $0.04^{+0.00}_{-0.02}$ & $0.05^{+0.00}_{-0.02}$ & $0.03^{+0.02}_{-0.00}$ \\
\hline
$\chi^2 / \text{d.o.f.}$ & 195 / 96 & 175 / 94 & 151 / 100 & 137 / 97 & 130 / 79 & 134 / 69 \\
\hline
\end{tabular}
~\\
$^f$ parameter frozen.
\end{center}
\end{table*}
\nopagebreak

\section{Discussion}

For GX~339--4, there is strong evidence that the central black hole has a spin parameter close to the maximal value of $a=0.998$ \citep[e.g.][]{Miller2008, Garcia2015, Parker2016}. During the high soft state, the inner disc could extend into the ISCO \citep{Plant2014}. Recently, \cite{Sridhar2020} found that the truncation radius remains low ($< 9r_{\rm g}$) throughout the bright intermediate state (i.e., the width of the Fe~K emission line is nearly constant throughout this transition indicating a quasi-static truncation radius). The evolution of the accretion disc during the intermediate state then should be driven by the variation in accretion rate rather than by the changes in truncation radius. The data anlyzed here captured the very last stages of the soft-to-hard transition (O1--O2) and the return to quiescence where the source was decreasing its luminosity (O3--O6).

\begin{figure}
    \centerline{
        \includegraphics[width=0.5\textwidth]{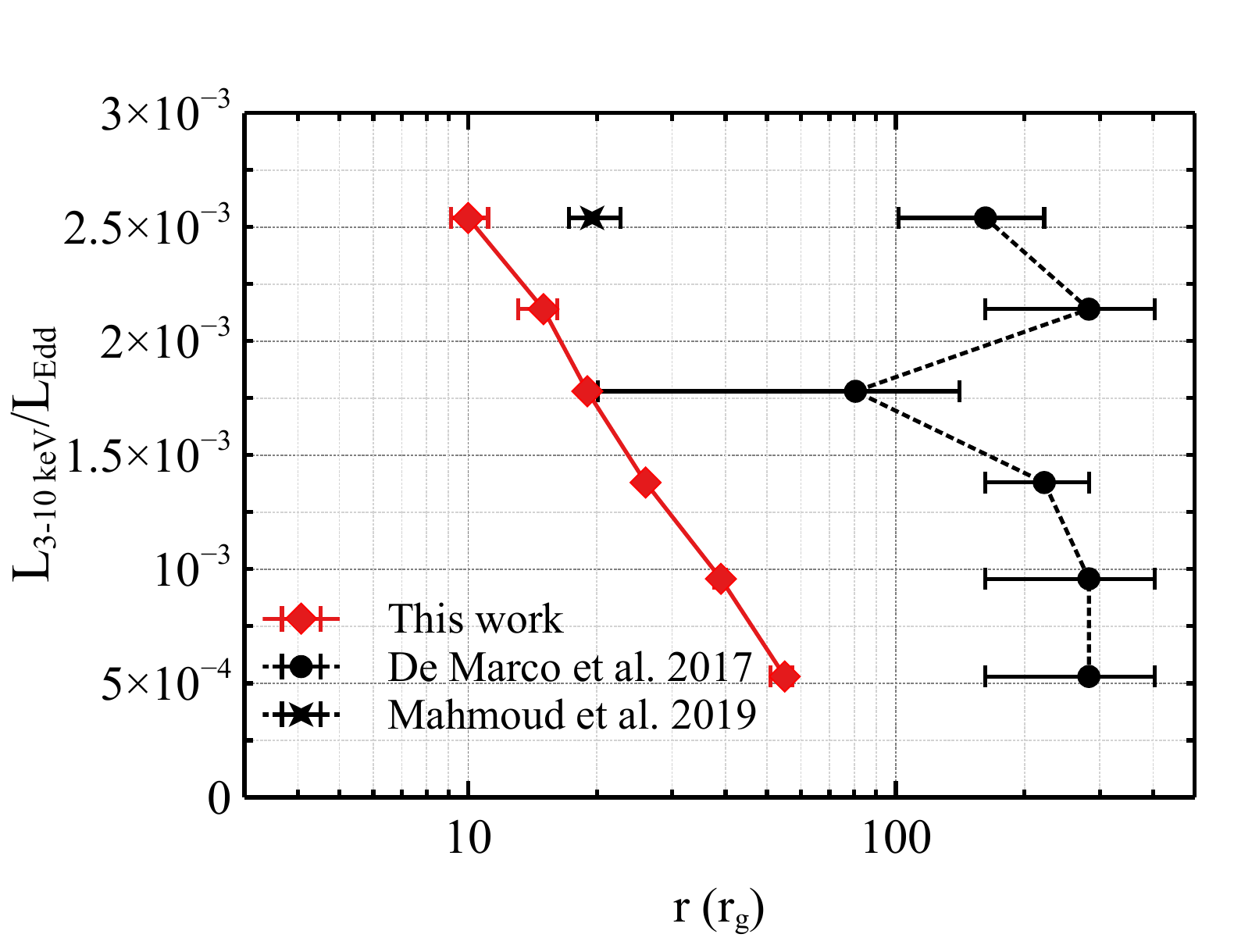}
    }
    \vspace{-0.2cm}
    \caption{Comparison of the truncation radius vs. Eddington-scaled luminosity for GX 339--4 inferred by this work (red diamonds), \cite{Demarco2017} via approximate reverberation measurement (circles) and \cite{Mahmoud2019} via spectral-timing fits (crosses).
    }
    \label{compare-rtrc}
\end{figure}

The data analysed here are similar to those used for the reverberation lag analysis of \cite{Demarco2017}, where they reported a decrease of the reverberation lag as a function of the source luminosity. Our results suggest that the truncation radius increases from $\sim 10 r_{\rm g}$ to $\sim 55 r_{\rm g}$ (see Table~\ref{tab_fit_para}) as the outburst proceeds and the flux decreases from O1--O6. The truncation radius versus the Eddington-scaled luminosity is shown in Fig~\ref{compare-rtrc}. The truncation radii from \cite{Demarco2017} are also shown in Fig~\ref{compare-rtrc}; these are estimated by converting lag amplitudes to the light-crossing distance from the centre to the inner edge of the disc, assuming a black hole mass of $10M_{\odot}$. The truncation radii being directly converted from the lag amplitude are significantly larger than those constrained by our PSD model. This is what we expected because the estimates by \cite{Demarco2017} are highly simplistic whereas our model accounts for the dilution. Our results support the commonly agreed framework that the disc truncation radius increases as the luminosity decreases \citep[e.g.][and references therein]{Wang2018}.    

\cite{Mahmoud2018} proposed a framework to explain the  time-averaged spectrum, the PSD in different energy bands and the time lags, simultaneously. Their model includes at least three distinct Compton components associated with three specific radii: the truncation radius, the inner radius of the hot flow and the jet-launch radius. Later on, \cite{Mahmoud2019} used their spectral-timing model, taking into account the effects of reverberation, to fit the O1 data, and found a truncation radius of $\sim 19.5 r_{\rm g}$. Their result is also plotted in Fig~\ref{compare-rtrc}. The truncation radius for O1 constrained by our PSD model is $\sim 10 r_{\rm g}$, which is smaller but still comparable to \cite{Mahmoud2019}. Spectral fits of the 2015 outburst data observed by {\it NuSTAR} and {\it Swift} were carried out by \cite{Wang2018}, where the truncation radius was found to be $\sim 3$--$15r_{\rm g}$. The accretion disc implied by the spectroscopic analysis \citep[e.g.,][]{Wang2018,Garcia2019} is likely to be truncated at smaller radii than all of those suggested by timing analysis.  

In principle, the truncation radius, $r_{\rm trc}$, and the X-ray source height cannot be well-constrained independently via reverberation lag measurements only since both of them directly determine the measured light-travel distance. Furthermore, from Table~\ref{tab_fit_para}, the frequency break index $\beta$ decreases as the outburst proceeds, suggesting that there is a variation in accretion rate as well as in $r_{\rm trc}$. Decreasing $\beta$ means the variability power in the frequency domain induced by the mass accretion rate fluctuations varies, from O1--O6, in a way that produces relatively high power at frequencies $f > f_{\rm break}$ (see eq.~\ref{eq:mdot_f}). Recently, \cite{Mushtukov2018} described the propagation of the fluctuations in the disc using the diffusion equation whose solution was derived by the method of Green functions. The suppression of variability at high frequencies was produced by the Green functions and was stronger for a lower parameter associated with radial kinematic viscosity. Our parameter $\beta$ that suppresses variability at high frequencies above the viscous frequencies then could describe dependence of kinematic viscosity and possibly tie back to the Green's function treatments such as in \cite{Mushtukov2018}. If the changes in luminosity are also driven by the variation in the viscous parameters, the frequency of $f_{\rm break}$ is changed as well (see eqs.~\ref{eq:fvis}--~\ref{eq:fbff}). We note that to limit the number of free parameters and avoid model degeneracy we fix $(H/R)^2 \alpha=0.005$ (i.e., assuming this parameter does not significantly vary during O1--O6). This value is obtained from the fits using the global grid and is in the acceptable range reported in the literature \citep[e.g.,][]{Veledina2016}.

\begin{figure}
    \centerline{
        \includegraphics[width=0.5\textwidth]{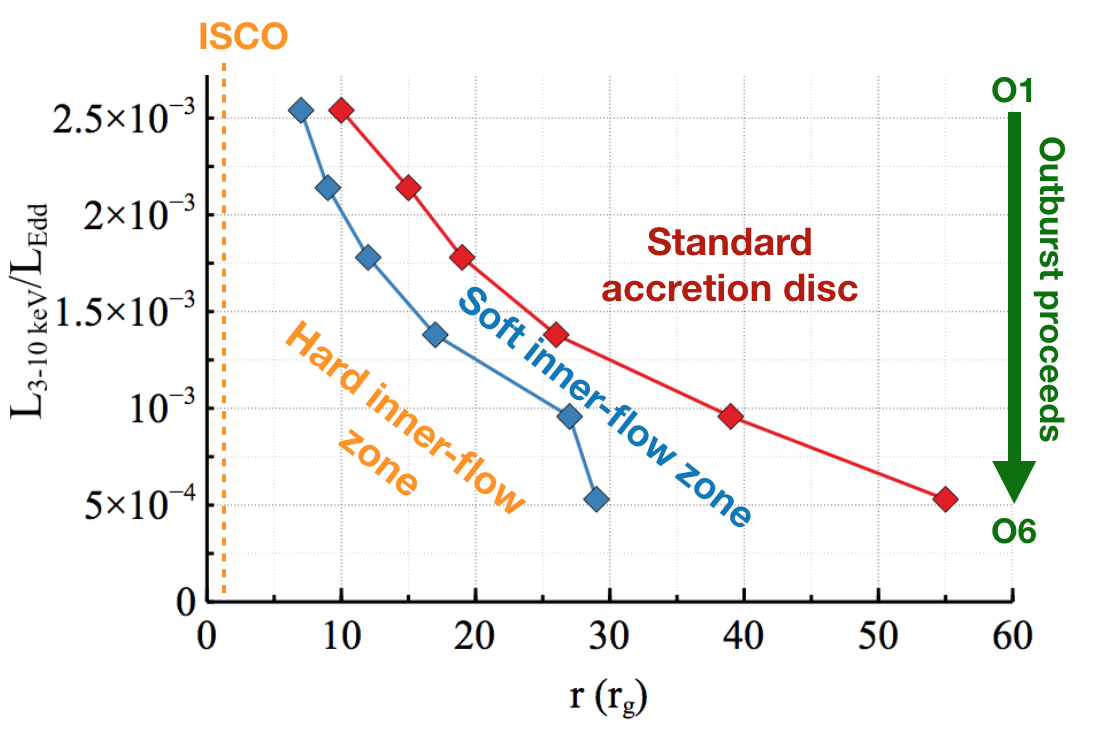}
    }
    \vspace{-0.2cm}
    \caption{The comparative size of inner-hot flow zones between different observations (O1--O6) implied from the best-fit parameters. Error bars are small and are excluded in this plot. 
    }
    \label{inner-flow-size}
\end{figure}

Furthermore, \cite{Mahmoud2019} fitted the spectral-timing data of O1 and reported the transition radius where the spectrally soft flow changes to hard inner flow to be $\sim 6.6 r_{\rm g}$. We estimate $r_\text{sh}$ for O1 to be $\sim 7 r_{\rm g}$, which is in agreement with their result. Starting from $r_\text{sh} = 7 r_{\rm g}$ at O1, our fitting also suggests that the soft-hard transition radius keeps increasing to $r_\text{sh} = 29 r_{\rm g}$ towards O6. The size of the spectrally soft inner-flow and hard inner-flow zones can be estimated as $r_{\rm trc} - r_{\rm sh}$ and $r_{\rm sz} - 1.235r_{\rm g}$, respectively. The comparative size of these flows is shown in Fig.~\ref{inner-flow-size}. As the flux decreases from O1--O6, the size of hard zone expands from $\sim 5r_{\rm g}$ to $27r_{\rm g}$, which is $\sim 1.1-2.2$ times the radial extent of the soft zone that expands from $\sim 3r_{\rm g}$ to $26r_{\rm g}$. 

Since we fixed $\Gamma_\text{sz}=2.0$ and $\Gamma_\text{hz}=1.5$ (consistent with the values reported by \cite{Demarco2017}), variations in $N_{F}$ (see Table~\ref{tab_fit_para}) could actually be due to the variations in spectral indices of the hot flows between different observations. The viscous timescale at the truncation radius, or the outer radius of the hot flow, sets the low-frequency break of the PSD. A low-frequency QPO due to Lense-Thirring precession of the hot flow \citep{Ingram2009} is sometimes observed in BHXBs in such a way that the QPO frequency moves with the low-frequency break. However, the QPO was previously found to be weak throughout O1--O6 \citep{Demarco2017, Mahmoud2019}, so they are not included in our fits. Additional Lorentzian or Gaussian components associating with these QPOs could definitely improve the fits, especially for the O1, but the trend of the PSD model and the implied parameters should not be significantly different.

Our model could allow the emissivity indices to be free but the exact values should depend on the assumed source geometry. Different geometries of the flow produce different illumination patterns on the disc and hence $\gamma$ could also vary among observations. However, we find that $\gamma \sim 0$ is obtained from the global fits in all six observations. It should be noted that $\gamma = 0$ produces same number of photons in each annulus of area $2\pi r {\rm d} r$ and that the emissivity in terms of flux per unit area on the disc is $r^{-1}$. This may suggest that the inner flows are significantly vertically extended in the way that the number of photons illuminating the outer part of the disc is large enough that make the emissivity profile likely flat. It may also require the inner flow to be more vertically extended than the outer flow so that some photons that are supposed to illuminate the thin disc near $r_{\rm trc}$ are obscured by the outer flow, flattening the emissivity especially at the inner parts of the disc. We caution that this is just a rough approximation and interpretation suggested by the global fits. In principle, $\gamma$ should not be thought of as an independent parameter but instead should tie to the exact geometry of the flow. Additional function may be required to tie $\gamma$ to characteristic radii of the flow. In that case, the exact geometry of the flow and also the disc (e.g., in case of flared disc) must be explicitly specified that makes the model much more complicated. We then select to fix $\gamma$ to the values obtained from the global fits in this work. Nevertheless, the disc emissivity profile is worth investigating in the future, using high density disc models \citep{Jiang2019}.

Dual lamppost sources illuminating the disc were found to be able to explain the spectroscopic data of GX~339--4 in some observations compared to the single lamppost source \citep{Garcia2019}. Therefore, the X-ray source could be extended and the realistic impulse responses need to be produced through the ray-tracing simulations  \citep{Wilkins2016, Chainakun2017, Chainakun2019b, Garcia2019}. Here, we assume the size scale of the source, or the hot-flows, to be relatively small compared to the radial size of the accretion disc \citep{Gardner2014, Mahmoud2019}. However, \cite{Chainakun2019a} reported that the reverberation signatures on the PSD profiles (e.g., the main dip and the oscillation features) would be relatively weak and become more difficult to probe if the source is significantly more extended. Using more realistic, ray-traced impulse responses from different spatially extended source geometries is beyond the scope of this paper, but is planned for the future.  

Although we consider two temperature inner flow plus thermal reverberation, the model can naturally produce the wiggles in the PSD profile which are usually seen in the observed data. The bumpy PSD could also be produced by other realistic treatments such as when the propagating fluctuations are solved through the diffusion equation using the Green function \citep{Mushtukov2018} and when extra variabilities are injected at different characteristic radii \citep[e.g.,][]{Rapisarda2016,Rapisarda2017}. Since the number of free parameters have been minimized to avoid degeneracies, it can be seen that the model provides very good constraint to each key parameter with small statistical errors. The power extracted from observational data towards the high-frequency end at some points turn out to be zero or negative, which are excluded in our analysis. New data delivered by upcoming X-ray observatories could provide higher quality data (e.g., mean spectrum, lags, and PSD) that could be robustly constrained with the model. High quality PSD extending to higher frequencies may reveal the clear dips and oscillatory structures produced by reverberation that are imprinted on the PSD. 

At the moment, we do not directly consider the effects of the response matrix of {\it XMM-Newton} that can misclassify photons at low energies \citep[see][and discussion therein]{Ingram2019}. This, however, could affect the timing data in the same way of changing the contamination flux in the thermal reverberation dominated and continuum dominated bands, which has been taken into account in this work by employing the reflected response fraction to determine the contamination ratio between these cross-components. Applying the response matrix directly to the model would be more straightforward and is worth investing in the future. Last but not least, our model can be further developed to predict the lag-frequency spectrum between these two bands. Fitting the PSD and the lags simultaneously (or the full cross-spectrum) is challenging and optimal statistics must be investigated to account for significantly different numbers of bins between different data set, otherwise the fitting might be biased towards the best-fit parameters for the PSD data that have significantly larger numbers of data points. A development of a more self-consistent model to simultaneously explain multi-timing data including the effects of the instrument response, however, is planned for the future.

\section{Conclusion}

In this study we develop a PSD model that is then fit to the data of GX~339--4 observed during the end of the 2015 outburst. We model a truncated accretion disc from $r_{\rm trc}$--$400r_{\rm g}$. Inside $r_{\rm trc}$, spectrally soft and hard hot-flows take place down to the ISCO. The model incorporates both disc-fluctuation and reverberation signals. The fluctuations inside the hot flows are propagate inwards on the viscous timescale of the truncation radius. The model can qualitatively reproduce the traditional PSD profiles exhibiting increased high-frequency power for higher energy band. Furthermore, stronger high-frequency power could be produced with decreasing $r_{\rm trc}$, which is also expected. Including the reverberation signals produces a dip at the high frequency end which should be the beginning of oscillatory structures imprinted in the PSD profiles as discussed by \cite{Papadakis2016} and \cite{Chainakun2019a}.

To produce the model grids, the reflected response fraction of 0.3--0.7 and 0.7--1.5 keV band are fixed at the values constrained by the time-averaged spectral analysis \citep{Demarco2017}, as well as the photon index of the X-ray continuum associated with the soft and hard hot-flow zones. In doing this, the PSD models are tied to the realistic parameters obtained from the spectral fitting and the contribution of continuum flux in each energy band as the dilution of reverberation is properly taken into account. The PSD data for both energy bands are fit simultaneously for each observation. We find the disc parameter $(H/R)^2 \alpha = 0.005$ could provide a good fit for the majority of observations using the global grid of the model, hence it is fixed when the finer, local grids are produced independently for each observation and the fitting is repeated. 

The fitting results suggest that the truncation radius moves outward from $10$--$55r_{\rm g}$ as the source luminosity decreases from O1--O6. Although the trend of increasing $r_{\rm trc}$ with decreasing luminosity is in agreement with previous studies, the values we obtain are smaller than previous reverberation lag analysis \citep{Demarco2017} and spectral-timing modelling \citep{Mahmoud2019}, but are larger than some of those constrained using spectroscopic data alone \citep[e.g.][]{Wang2018,Garcia2019}. We find that the transition radius also increases from $7$--$29r_{\rm g}$ during O1--O6, meaning that the size of inner hard hot-flows increase from $\sim 5$--$27r_{\rm g}$, which always span a slightly larger radial range than the spectrally soft hot-flows by a factor of $\sim 1.1$--$2.2$. The current PSD model can be straightforwardly adapted for different source and inner-flow geometries, e.g., dual lamppost cases, that may suit the unique data of different X-ray binaries and AGNs.

\begin{acknowledgements}
The calculations in this work were carried out using the BlueCrystal supercomputer of the Advanced Computing Research Centre, University of Bristol, UK. All data analyzed are based on observations obtained with {\it XMM-Newton}, an ESA science mission with instruments and contributions directly funded by ESA Member States and NASA. PC thanks the Thailand Research Fund (TRF) for support under grant number MRG6280086, and acknowledges useful discussions with Utane Sawangwit. WL thanks for the financial support from the Faculty of Science, Srinakharinwirot University. We thank the anonymous referee for their comments which led to a clarification of some important points in the paper and has inspired some interesting work for the future.
\end{acknowledgements}

%
%

\end{document}